# Identity-Based Attribute Prototypes Distinguish Communities on Twitter

Thomas Magelinski and Kathleen M. Carley

Software and Societal Systems Department, Carnegie Mellon University, 5000 Forbes Avenue, Pittsburgh, 15213, PA, USA.

*Corresponding author(s). E-mail(s): tmagelin@cs.cmu.edu;

**Abstract**

This paper examines the link between conversational communities on Twitter and their members' expressions of social identity. It specifically tests the presence of community prototypes, or collections of attributes which define a group through meta-contrast: high in-group cohesiveness and high out-group distinctiveness. Analyzing four datasets of political discussions ranging from roughly 4 to 30 million tweets, we find strong evidence for the presence of distinctive community prototypes. We observe that community prototypes are constructed through hashtags, mentions, emojis, and identity-phrases. This finding situates prior work on the identity signaling of individual users within a larger group process playing out within communication communities. Community prototypes are then constructed for specific communities by measuring the salience of identity signals for each community. Observed community prototypes tend to be based on political ideology, location and language, or general interests. While the presence of community prototypes may be a natural group behavior, the high levels of contrast observed between communities displaying ideologically opposed prototypes indicate the presence of identity-related polarization.

## Introduction

Group-level processes are at the heart of many pressing problems on social media such as polarization [30, 43], radicalization [12, 40], and the diffusion of misinformation [10, 17, 41]. For offline networks, the social identity perspective has been taken to make headway on these problems. Social identity theory and self-categorization theory have been validated and used to understand intergroup conflict in organizations [24], Islamic Extremism [1], American political polarization [23, 27, 31], and hostile media perceptions [49, 52]. These successes suggest that the social identity perspective has great potential for understanding these pressing issues in the social media setting. However, the social identity perspective relies on an understanding of the relationship between individuals' social identity and the communities they are a part of. While prior work has found that social media users signal their social identity, the connection to online communities has thus far been unclear.

The core idea of the social identity perspective is that people construct their self-concept in part from the communities and categories that they belong to [26, 54]. Group-level processes then arise from individuals' social cognitive processes based on their social identities. Under self-categorization theory, social identities can be understood through community prototypes, or fuzzy sets of attributes which define a group and distinguish it from others [29, 59]. Attribute's maximization of in-group similarity and out-group difference is known as the meta-contrast principle. Individuals' feelings about themselves and others are functions of their alignment with the group prototype. Thus, developing an understanding of community prototypes on social media is crucial to the application of social identity-related theories.

Researchers have observed multiple mechanisms that social media users leverage to project their social identity. Hashtags are perhaps the most popular tool, enabling users to signal parts of their social identity and community membership in a searchable way [11, 25, 53]. Their searchable nature allows users to find others aligned with their social identity, and form online communities [18, 62]. Beyond hashtags, Twitter users have been observed to signal their social identity in their profile descriptions using "personal identifiers" or phrases that refer to an individual's social group, category, or role [46, 51, 63]. Users also commonly use emojis to indicate their political beliefs and interests [20, 22, 33, 36]. Subtleties in emoji usage such as emoji sequencing and the use of skin color modifiers have been observed to signal user identity with greater granularity [16, 50].

While prior works have observed social media users signaling social identity through a variety of mechanisms, they have not tied these identity attributes to specific online communities, short of communities that are themselves defined through the use of a single hashtag. These works could, for example, identify users describing themselves as "mothers", however, they cannot identify whether or not the social attribute of "mother" is salient for some observed community. Meaning they cannot say if there is a community of mothers or if there happen to be mothers in communities that are divided based on other attributes. In other words, identity attributes have been used to understand individual users, but have not been used to understand communities at scale.

The missing link has been a mechanism for quantifying an identity-based attribute's meta-contrast, or its ability to distinguish a particular community. We develop a multi-view network methodology to tackle this problem. First, we propose using multi-view projected modularity to quantify the *overall* strength of prototypes in the dataset. A low modularity value indicates poor separation between user communities based on their attributes, implying that prototypes are not present. On the other hand, a high value indicates that attributes strongly separate communities and a prototype is present. Then, we develop modularity vitality for bipartite projections in order to quantify an *individual attribute's* meta-contrast for each community. For a given community, the set of attributes with the highest values are considered the community's prototype, because those are the attributes which maximally contribute to the community's in-group similarity and out-group differentiation. The multi-view approach enables the detection of prototypes across the different modalities that users have been observed to signal their social identity such as putting hashtags, emojis, and personal identifiers in their biography. The multi-view approach enables the study of the different mechanisms or modalities of identity signaling seen in prior work.

Communities can be defined in several ways. Here, we are interested in conversational communities, or groups of users who are more frequently engaged in conversation with each other than with users outside of their community. Following communities, or communities based on following-relationships could also be studied, and there is evidence that these communities hold similar interests and beliefs [6, 32, 36, 37, 64]. However, these communities are roughly static compared to conversational communities, making the salience of attributes difficult to demonstrate. Further, conversational communities are commonly studied when measuring polarization and information diffusion making them most relevant to study for future applications [8, 21, 42, 61].

# Methods and Materials

## Data

We measure the presence of community prototypes for conversational communities on four Twitter datasets, each surrounding a different conversation. The conversations were selected to cover different aspects of political communication; the *Election* and *Reopen* datasets are explicitly political, while the *COVID* and *Captain Marvel* datasets are not, though they have political elements. The *COVID* dataset captures discussion surrounding the coronavirus pandemic starting from March 11, 2020, the day when WHO declared the coronavirus a pandemic[1], continuing for one week. The *Election* dataset captures discussion surrounding the 2020 US Presidential election starting from the day before the election November 2, 2020 and ending the day after major news outlets called the election, November 8, 2020. The *Reopen* dataset captures discussion surrounding the Reopen America protests starting from April 1, 2020 and ending on June 22, 2020. The *Captain Marvel* dataset captures discussion surrounding the Captain Marvel movie, Marvel's first movie with a female lead, from February 15 to March 15, 2019. Each dataset was collected using keyword-based searches, as is typical in the study of social media data [8, 21, 22, 42, 53]. The keywords for each dataset can be found in the Appendix. For each dataset, the communication network was derived, and the Leiden algorithm was used to uncover communities [57]. Network statistics for each dataset can be found in Table 1.

[1] https://pubmed.ncbi.nlm.nih.gov/32191675/

| Dataset | Tweets | Users | Edges |
|---|---|---|---|
| Reopen | 10,131,537 | 3,495,506 | 11,032,399 |
| Election | 4,248,125 | 1,814,513 | 7,611,473 |
| COVID | 29,498,233 | 9,888,775 | 35,288,357 |
| Captain Marvel | 5,455,142 | 1,642,434 | 4,981,094 |

**Table 1**: Basic networks statistics for each dataset.

## Prototype Measurement and Construction

### Prototype Measurement with Projected Modularity

To measure the presence of community prototypes, we begin by modeling the user-attribute relationship as a multi-view network. Each view of the network corresponds to a different attribute-type. Based on the literature on Twitter users' ability to signal social identity through multiple modalities, we consider 6 attribute types. From a user's free-text biography we consider hashtags, mentions, personal identifiers [46], and emojis. We also extract hashtags within a user's name, and unigrams in their location field. For each attribute type, a user-attribute bipartite network is constructed, where users are connected to the attributes they exhibit. Each bipartite view is projected onto the user nodeset, such that a user-to-user network is obtained. Connections in these views indicate the number of attributes that a pair of users have in common.

More formally, we begin with a bipartite network $G$. This network connects users to the attributes that the exhibit, only considering a single "attribute-type" for now. The information from this network is encoded in the adjacency matrix, $B$, where $B_{i,j} = 1$ if user $i$ exhibits attribute $j$ and is 0 otherwise. We fold this network to obtain a user-to-user network where edges are weighted by the number of attributes that the two users have in common. This network's information is encoded in an adjacency matrix, $A = BB^T$. Examining the folded network allows us to study the relationship between attributes and user communities without clustering the attributes. This process is repeated for each attribute type, $v$, which results in a multi-view network which each attribute type's view encoded in an adjacency matrix, $A_v$.

This framework enables us to quantify the presence of prototypes with the well-known network measure, modularity. More specifically, modularity in this case quantifies the *meta-contrast* exhibited by the communities. Under self-categorization theory, community prototypes are constructed with attributes maximizing *meta-contrast*, referring to the dual goal of simultaneously maximizing in-group similarity and minimizing out-group similarity [59]. The higher the meta-contrast, the stronger the prototypes.

From a network perspective, in-group similarity can be quantified by the number of shared-attribute connections within a community, known as internal edges. Internal edges represent pairs of same-community users sharing an attribute; the more of them there are, the higher the group's cohesiveness. Given a vector of node communities $\mathbf{c}$, where $c_i$ indicates the community of node $i$, the number of internal edges in the network is given by $\frac{1}{2}\sum_{i,j} A_{i,j}\delta(c_i, c_j)$, where $\delta$ is an indicator function equaling one if the two arguments are equal and equaling zero otherwise. Similarly, out-group similarity can be quantified by the number of edges falling between communities, known as external edges.

The balance between internal and external edges can be captured by using the fraction of internal edges. For the fraction to be high, there must be many internal edges *and* few external edges. Thus, the the higher the fraction, the larger the meta-contrast, and the stronger the evidence for prototypes. Figure 1 illustrates this. It shows that user communities which are well-separated by attributes will have a high fraction of internal links in the projected networks, while user communities which are poorly separated will have a low internal link fraction.

Modularity was developed to quantify the fraction of internal edges appearing in a network while accounting for those that would be expect by chance under a null model [44, 45]. The most common form of modularity is Newman Modularity, which considers a unipartite network (not a projection) and using the configuration model as a null model. Though multiple instantiations of the configuration model exist, the attribute networks studied here are considered "simple" where these differences are inconsequential [14]. Barber adapted this to bipartite networks, and Arthur built on this adaptation to develop a modularity for bipartite projections, which is given in Equation 1 [2, 5]. The set of communities is given by $\mathfrak{C}$, $F$ is the number of bipartite edges $F = \sum_{i,j} B_{i,j}$, $M$ is the sum of weighted projected edges $M = \frac{1}{2}\sum_{i,j} A_{i,j}$, the sum of weighted internal edges for a given community is calculated as $M_c^{int} = \sum_{i,j} A_{i,j}\delta(c, c_i)\delta(c, c_j)$, and strength of a community is given by $l_c = \sum_{i,j} B_{i,j}\delta(c, c_i)$.

$$Q^P(G, \mathfrak{C}) = \sum_{c \in \mathfrak{C}} \underbrace{\left( \frac{M_c^{int}}{M} - \left( \frac{l_c}{F} \right)^2 \right)}_{Q_c^P} \tag{1}$$

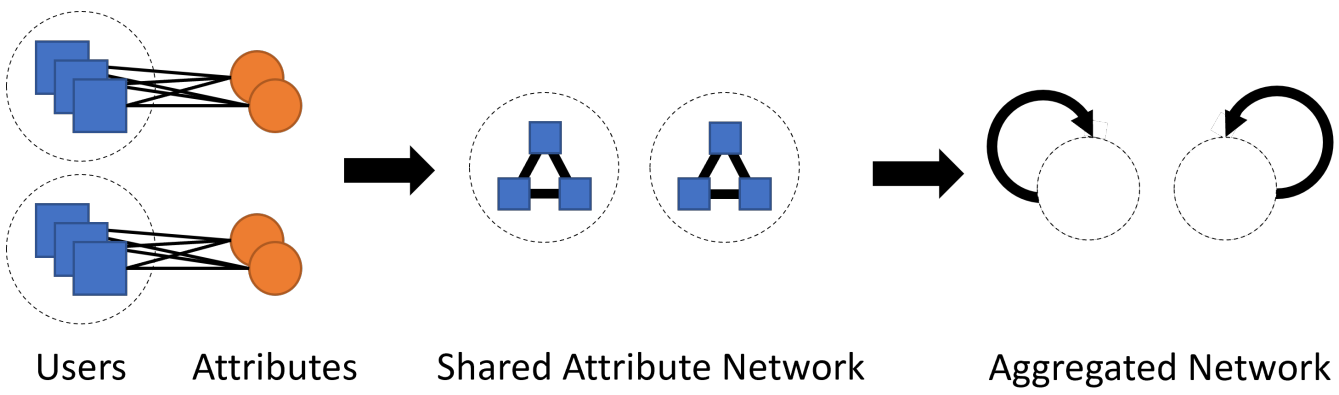

(a) Attributes distinguish communities.

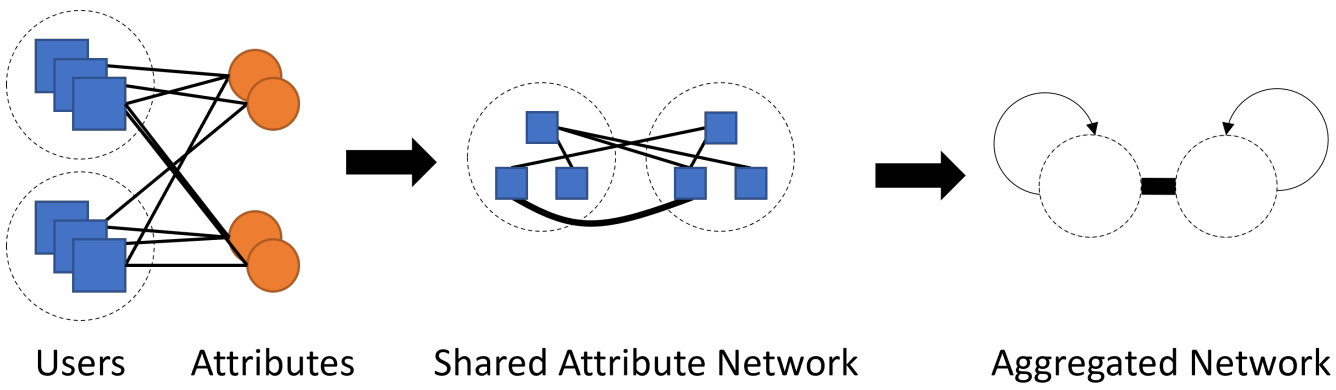

(b) Attributes shared across communities.

**Fig. 1**: An illustration that modularity of shared-attribute networks is indicative of how well-separated user communities are by their attributes. Users are represented by blue squares while attributes are represented by orange circles. The dotted circle around clumped squares indicates that they belong to the same community; there are two user communities in each example. The black arrow indicates the process of projecting the bipartite user-attribute network into a user-to-user network based on shared attributes. When attributes distinguish communities, as in Figure 1a, the communities are still well-defined in the projected shared-attribute network. This can be seen by the strong links within the community boundaries in Figure 1a, and the absence of links between communities. Figure 1b illustrates the case where attributes are shared across communities. This is easily seen in the shared-attribute network, where there are many inter-community edges and few intra-community edges. Projected modularity can quantify this phenomenon, resulting in a perfect score for Figure 1a, and a low score for Figure 1b.

A high value of $Q^P$ means that a high fraction of edges in the projected network are falling within the communities, after accounting for those which would fall there by chance under the bipartite configuration model [2].. Applying this to the user-attribute network, a high $Q^P$ indicates that many more of the shared-attribute relationships are happening between users that are in the same community than we would expect by chance. Thus, high values of $Q^P$ are indicative of communities exhibiting prototypes.

We calculate $Q^P$ for each user-attribute network using the communities obtained by community detection in the *interaction network*. The value observed for each user-attribute network indicates the extent to which prototypes are observed using that type of attribute. Along with classic interpretation of modularity, we consider values above 0.3 to give moderate evidence for the observation or prototypes, and values above 0.5 to give strong evidence [45]. We note that we are *not* performing modularity maximization in this step, as the communities have been derived separately on data that does not explicitly include any attributes.

### Community-Level Visualizations

To visually depict the modularity values, we plot the expected number of shared attributes from members of different communities (subtracted those expected due to chance) for the top 20 communities of all datasets in Figure 3. The process to construct these diagrams is as follows.

For every attribute type, a community-to-community shared attribute network was constructed, after filtering 2% of the non-salient attributes according to the Modularity Filtering method discussed later in this Section. The adjacency matrix, $A_v$ for each view $v$ of the network was calculated as $A_v = (C^T B_v)^T (C^T B_v)$, where $C$ is the user-community indicator matrix ($C_{i,j} = 1$ when user $i$ belongs to community $j$ and is 0 otherwise), and where $B_v$ is the filtered user-attribute bipartite adjacency matrix for attribute type $v$ ($B_{v,i,j} = 1$ when user $i$ exhibits attribute $j$ and is 0 otherwise).

Next, the expected number of shared attributes across communities under the configuration null model was calculated for each view. The expected adjacency matrix, $\mathbb{E}\|A_v\|$, was calculated as $\mathbb{E}\|A_v\| = 2M_v L_v^T L_v / F_v$, where $L_v$ is the vector of community strengths $l_c$ for view $v$ as previously defined, and $M_v$ and $F_v$ are the sum of edge weights in for view $v$ in the projected network and the bipartite network, respectively. This is the same number of expected internal links as used in Equation 1.

Then, the total number of shared-attributes between communities, above which is expected by chance, was calculated as: $\mathcal{A} = \sum_{v \in \mathcal{V}} A_v - \mathbb{E}\|A_v\|$. This collapses the views of the network, while accounting for the differing degree distributions of each of the views.

Lastly, the total number of shared-attributes between communities was normalized to indicate the number of expected shared-attributes (above chance) between any pair of users in the two communities. This is accomplished by divided the number of shared-attributes by the number of users in community 1 times that of community 2. The number of users in community $c$, $N_c$, can be calculated as $\sum_i C_{i,j}$. The normalization matrix is then $NN^T$. This normalization is applied element-wise to obtain the final adjacency matrix: $\bar{\mathcal{A}} = \mathcal{A} \oslash (NN^T)$. To be clear, the symbol $\oslash$ indicates element-wise division, such that $\bar{\mathcal{A}}_{i,j} = \mathcal{A}_{i,j} / (NN^T)_{i,j}$. The community-to-community visualization is finally drawn using edge weights corresponding to the entries in $\bar{\mathcal{A}}$. Edge weights below zero, indicating that there are *less* shared attributes between users of the communities than expected by chance, are not drawn.

## Prototype Construction with Projected Modularity Vitality

Prototypical attributes are defined to be those which maximize meta-contrast; that is, attributes that help define the group and differentiate it from others. In network terms, prototypical attributes are those which maximally contribute to community structure. We have quantified the overall level of attributes association with group structure using projected modularity. Now, we develop projected modularity vitality to quantify the contribution of individual attributes to community structure. Sorting attributes by this value then gives prototypical attributes, or those which are signaled by many members within a community and few outside of it.

Network vitalities, or induced centrality measures, are used to quantify a nodes contribution to a global network value [13, 34]. Previously, we developed modularity vitality to quantify node contribution to community structure in the unimodal case [39]. Now, we develop projected modularity vitality to quantify how much a node in the projected nodeset of a bipartite network contributes to communities in the opposing nodeset. Applied to our data, this will measure each attribute's contribution to each user community.

Network vitalities simply compare the original global network value to what it would be if a node and all its associated edges are removed from a network, as shown in Equation 2 where $G$ is the network, $i$ is the node to be removed, $F$ is the function giving the quantity of interest and $G - \{i\}$ is the network with node $i$ and its associated edges removed.

$$V_F(G, i) = F(G) - F(G - \{i\}) \tag{2}$$

We select $F$ in equation 2 to be $Q^P$ from equation 1. Similar to [39], we need to derive a computationally efficient form of Projected Modularity Vitality in order to apply it to large real-world network data. We do so by simply recognizing the impact of removing a node on the four terms in $Q^P_c$.

First, we define an attribute's degree in the bipartite network as $d_j = \sum_i B_{i,j}$, where $j$ indicates the attribute of interest. The total number of edges in the bipartite network goes from $F$ to $F - d_j$ when node $j$ is removed. We also define the community degree, $d_{j,c} = \sum_i B_{i,j}\delta(c, c_i)$, which gives the number of users in community $c$ displaying the attribute of the network. The strength of community $l_c$ becomes $l_c - d_{j,c}$ when node $j$ is removed. A property of network projection is that a node with degree $d_j$ will yield edges whose weights sum to $\frac{1}{2}d_j^2$ in the projection. Thus, $M$ becomes $M - \frac{1}{2}d_j^2$ after the removal of node $j$. Similarly, $M_c^{int}$ becomes $M_c^{int} - \frac{1}{2}d_{j,c}^2$. These results give the equation for projected modularity vitality and its computation in Equations 3 and 4, respectively.

$$V_{Q^P}(G, \mathfrak{C}, j) = Q^P(G, \mathfrak{C}) - Q^P(G - \{j\}, \mathfrak{C} - \{j\}) \tag{3}$$

$$Q^P(G-\{j\},\mathfrak{C}-\{j\}) = \sum_{c\in\mathfrak{C}}\left(\frac{M_c^{int}-\frac{1}{2}d_{j,c}^2}{M-\frac{1}{2}d_j^2} - \left(\frac{l_c-d_{j,c}}{F-d_j}\right)^2\right) \quad (4)$$

$$V_{Q_c^P}(G,\mathfrak{C},j) = \left(\frac{M_c^{int}}{M} - \left(\frac{l_c}{F}\right)^2\right) - \left(\frac{M_c^{int}-\frac{1}{2}d_{j,c}^2}{M-\frac{1}{2}d_j^2} - \left(\frac{l_c-d_{j,c}}{F-d_j}\right)^2\right) \quad (5)$$

Finally, we note that projected modularity vitality is naturally broken up into community terms, allowing for the quantification of a node's contribution to each community individually. This contribution, for a node $j$ is given in Equation 5. For each community, the terms with the highest values of $V_{Q_c^P}$ are taken to be prototypical.

The modularity vitality approach identifies attributes which are mostly exhibited by a single community, and which are popular within that community. This is a necessary improvement over, for example, relative frequencies, which are likely to identify less common attributes.

## Modularity Filtering

We previously stated that if communities exhibit prototypes, "members within a community will share a set of attributes with each other and they will not share these attributes with other communities." We note that it is still possible for a set of non-prototypical attributes to be shared among members of all communities.

Consider the social circles on a college campus. Each social circle has its own set of prototypical attributes, yet all people involved share the attribute that they are a student of the same college. This is not a *salient* attribute in the present definition of communities, so it does not affect whether or not prototypes are present. However, under our modularity framework, the inclusion of the college attribute would decrease the fraction of internal edges and thus lower the perceived strength of prototypes. If many of non-salient attributes are present, prototypes may be effectively drowned out. An illustration of the effect is given in Figure 2.

We can identify non-salient attributes as those with negative modularity vitality scores, which are known as "community-bridges" because they create shared-attribute edges between users of differing communities. Removing the top non-salient attributes is akin to the modularity filtering approach applied in [38, 39] or the modularity-vitality backbone approach in [48]. It can also be seen as an "initial" network attack when viewed from a network robustness perspective [28].

We apply this procedure to the top 2% of non-salient nodes to uncover a more accurate measurement of the presence of prototypes in the data. We have shown the top 5 attributes removed in each category, as well as the top 5 *most* salient attributes in each category. Unlike the previous analysis, this is a type of modularity-maximization procedure, so needs to be statistically tested against a null model. Thus, we used the bipartite configuration model

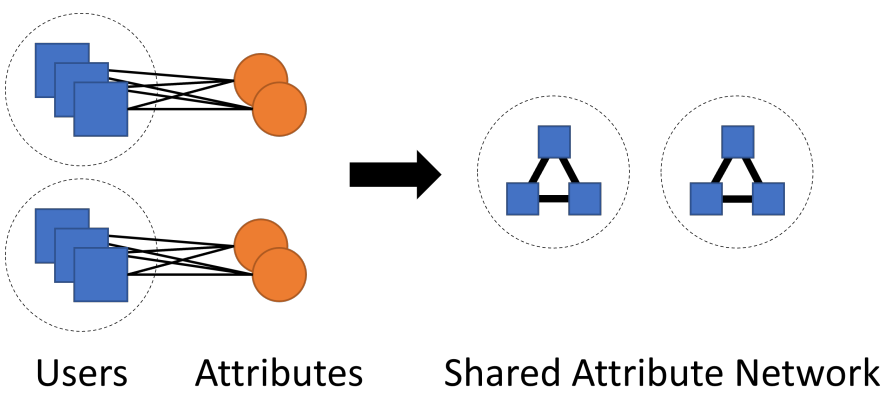

(a) Original network.

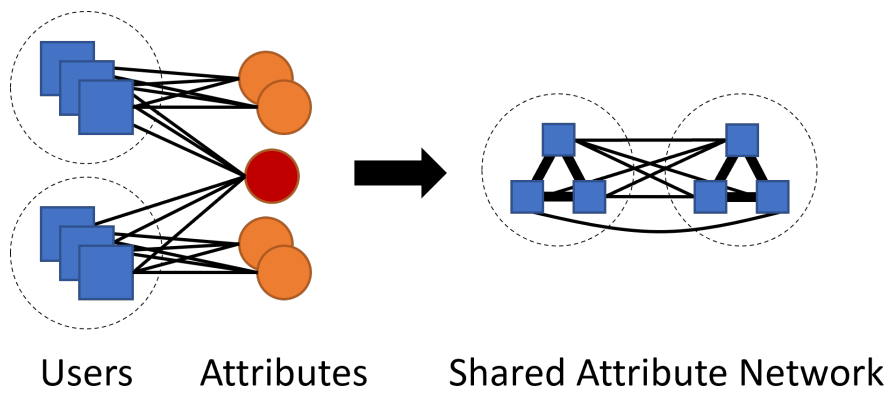

(b) Original network with the addition of a popular but non-salient attribute.

**Fig. 2**: An illustration that the addition of a non-salient attribute can lower the perceived modularity. In Figure 2a, the communities are well-separated by attributes. In Figure 2b, an additional attribute has been introduced (red circle). This attribute is shared by all members, so is non-salient, however results in many external connections in the projected network, thereby lowering the modularity.

as our null model using the real degree sequences observed in our dataset and performed filtering on each one of these randomly generated networks. The process was completed 250 times to provide confidence similar to that of using p-values at the value of $p < 0.004$. The filtered modularity value on the real data was compared to the distribution of scores obtained from the generated networks.

## Multi-Modal Analysis

The previous methods detail our approach for a single attribute type. We extend this to the analysis of multiple attribute types through multi-view modularity [9]. Under this framework, each attribute type creates a "view" in a multi-view user-to-user shared attribute network. Multi-view modularity simply takes a weighted average of the modularities of individual views: $Q = \frac{1}{\sum_{v\in\mathcal{V}} w_v} \sum_{v\in\mathcal{V}} w_v Q^v$, where $\mathcal{V}$ is the set of views, $Q^v$ is the modularity for view $v$ and $w_v$ is the weight of view $v$. The weights may be manually set to enforce the importance of some views over others. In our case we equally weight views with $w_v = 1$. Using this framework, quantifying the contribution of attributes across attribute types is now possible and attributes can be directly compared across views. To make values more comparable across communities, they are normalized according to their multi-view modularity, which is just the average of the modularity values across all views

$$\overline{MV}_{c,j} = \frac{\|\mathbb{V}\|}{\sum_{v\in\mathbb{V}} Q_c^{P,v}} V_{Q^P_{c,j}}, \tag{6}$$

where $\mathbb{V}$ is the set of attribute views.

### Computation

While mathematically we are operating on the projected shared-attribute networks, this is often not computationally feasible. Though real-world networks tend to be sparse, this is not necessarily the case with real-world projected networks. Consider an example where 1 million users share a single attribute. The bipartite network only has 1 million links, whereas the projected network will be fully-connected and contain 1 trillion links. The *COVID* dataset is the largest dataset we examine, making it the most computationally expensive to analyze. The sum of projected edge weights ranges from roughly 6.5 million (name hashtags) to to 167 billion (location unigrams).

Thus, all computation is performed on underlying bipartite networks. This is made possible by the fundamental property of projection: an attribute exhibited by $d$ users (attribute has degree $d$ in the bipartite network), will result in $\frac{1}{2}d^2$ in the projected network. This property allows for the computation of $M$ and $M_c^{int}$ without the construction of the projected network: $M = \sum_j \frac{1}{2}d_j^2$ and $M_c^{int} = \sum_j \frac{1}{2}d_{j,c}^2$. This trick makes the computation of $Q^P$ feasible even in the case that the projection yields an extremely large and dense network. Computation of modularity vitality is even more costly than modularity. Thus, the same computational trick is used and all calculations are made on the underlying bipartite networks. Refer to the accompanying code for full details[2].

## Results

### The Presence of Prototypes

Results for each dataset are presented in Table 2. First, we consider the raw, or unfiltered data in column 3. Across all datasets, we see that user communities are strongly separated by attributes, at least across some attribute types. Hashtags and mentions in user biographies are the strongest indicators of community, with modularity values ranging from roughly 0.18 to 0.63. Hashtags in usernames are also strong indicators of community. Personal identifiers, emojis, and location unigrams have low to moderate values.

The surprisingly low modularity values of signals like bio personal identifiers and location unigrams can be explained and accounted for by considering the salience of attributes. Personal identifiers and location unigrams are free-text attributes, naturally resulting in a less unified presentation of attributes and creates far more non-salient attributes than attributes like hashtags, which have a mechanism for seeing people and content using the same exact indicator as you. Thus, we consider the results on the filtered network, given in the fourth column of Table 2.

Now, personal identifiers have reasonably high modularity values, around 0.2 for all datasets except *Captain Marvel*. We also see large gains in the location unigrams attributes. A statistical test of the results was performed by performing the procedure on network generated from the configuration null-model 250 times. All results were found to be significant to $p <= 0.004$. Overall, we see even stronger evidence that conversational communities differentiate themselves via multi-modal prototypes.

[2]Prototype construction of Twitter conversation communities has been implemented in the ORA-Pro network analysis software https://netanomics.com/ora-pro/. The original code for this work will be made available at https://github.com/tmagelinski

| Dataset | Attribute | Modularity | 2% Filtered Modularity |
|---|---|---|---|
| Reopen | Bio Personal Identifiers | 0.1717 | 0.2655 |
| | Bio Mentions | 0.3794 | 0.6862 |
| | Bio Hashtags | 0.5655 | 0.7168 |
| | Bio Emojis | 0.1011 | 0.1768 |
| | Name Hashtags | 0.4299 | 0.5132 |
| | Location Unigrams | 0.0859 | 0.2426 |
| Election | Bio Personal Identifiers | 0.0795 | 0.2294 |
| | Bio Mentions | 0.3323 | 0.4509 |
| | Bio Hashtags | 0.2909 | 0.3987 |
| | Bio Emojis | 0.1385 | 0.2116 |
| | Name Hashtags | 0.1008 | 0.2216 |
| | Location Unigrams | 0.0885 | 0.1607 |
| COVID | Bio Personal Identifiers | 0.1326 | 0.1988 |
| | Bio Mentions | 0.3648 | 0.6981 |
| | Bio Hashtags | 0.6304 | 0.7631 |
| | Bio Emojis | 0.0368 | 0.0796 |
| | Name Hashtags | 0.5860 | 0.7476 |
| | Location Unigrams | 0.2770 | 0.5633 |
| Captain Marvel | Bio Personal Identifiers | 0.0522 | 0.0863 |
| | Bio Mentions | 0.1889 | 0.4385 |
| | Bio Hashtags | 0.3542 | 0.5755 |
| | Bio Emojis | 0.0173 | 0.0346 |
| | Name Hashtags | 0.3005 | 0.3470 |
| | Location Unigrams | 0.0562 | 0.2301 |

**Table 2**: Projected Modularity Values for each dataset. Filtered Modularity values were found to be significant $p <= 0.004$

To understand the attributes that were filtered, the most and least salient (highest and lowest modularity vitality) attributes of each type for the *Election* dataset are given in Tables 3 and 4. Tables for the other datasets are given in the Appendix. In this dataset, personal pronouns were not salient overall. We also see that affiliation or support of different soccer teams was not salient. Neither were were less-politically charged emojis like the heart emojis. Attributes associated with support for Donald Trump are salient, including personal identifiers like maga and patriot, hashtags like #maga, and the American flag emoji, 🇺🇸. Politically liberal attributes like #resist, and #blacklives matter, are not often not salient. This could indicate that Trump-supporting users are isolated in a small number of communities while Biden-supporting or otherwise left-leaning users are dispersed in many conversational communities.

We observe that, paradoxically, #bidenharris2020 is one of the *most* salient hashtags when displayed within a name, but one of the *least* when displayed in a biography. This is counterintuitive but could be indicative of subtle usage differences across sub-communities. This outcome is consistent with a scenario

| Personal ID | | Mention | | Hashtag | | Emoji | |
|---|---|---|---|---|---|---|---|
| **S** | **NS** | **S** | **NS** | **S** | **NS** | **S** | **NS** |
| maga | she | @genflynn | @manutd | #maga | #blacklivesmatter | 🇺🇸 | ❤️ |
| patriot | her | @realdonaldtrump | @arsenal | #kag | #blm | 🌊 | 💙 |
| conservative | he | @potus | @bts_twt | #fbpe | #resist | 🇮🇳 | 🏳️‍🌈 |
| christian | him | @joebiden | @chelseafc | #trump2020 | #bidenharris2020 | 🇫🇷 | ✨ |
| wife | writer | @kamalaharris | @lfc | #noafd | #fbr | 🇬🇧 | 🤍 |

**Table 3**: The most salient (S) and least salient (NS) attributes of each attribute derived from user biographies within the *Election* Dataset

| Name Hashtag | | Location Unigram | |
|---|---|---|---|
| **S** | **NS** | **S** | **NS** |
| #fbpe | #blm | usa | england |
| #maga | #endsars | france | new |
| #bidenharris2020 | #blacklivesmatter | india | ca |
| #stopthesteal | #biden | brasil | the |
| #trump2020 | #biden2020 | venezuela | united |

**Table 4**: The most salient (S) and least salient (NS) attributes of each attribute *not* derived from user biographies within the *Election* Dataset

where a hashtag is widely used in a biography, but only a specific community of users puts it in their name. Given that putting a hashtag in your name is a more prominent display than in your bio, it is plausible that these more extreme users would be more concentrated in communities and less dispersed throughout the network.

Hashtags in favor of Black Lives Matter or the Democratic Party are typically not salient in the overall network. A possible explanation for this is that there are pro-Democrat conversational communities. Under this scenario, attributes like #resist will form many cross-cutting connections between these aligned communities. On the other hand, the overall salience of pro-Trump and pro-Republican attributes suggests that Trump supporters are concentrated in fewer conversational communities. We see that this is the case in the following section studying the prototypes of individual communities.

Lastly, the attribute-projection procedure illustrated in Figure 1 was carried out on the top 20 communities in each of the datasets, ranked by their contribution to modularity, $Q_c^P$. For the *Election* dataset, these communities contain 83.9% of the users. The coverage in other datasets is similar. After filtering the same 2% of attributes, the result was visualized in Figure 3. The resulting corresponding diagram for the unfiltered data is shown in the Appendix. We make four observations. First, the strong within-community edges (or loops) and the thin between-community edges illustrate the presence of prototypes for all datasets. Second, we see that prototype strength varies by community. In the *Reopen* dataset, for example, community 10 has a strong prototype, while community 14 does not. Third, some community prototypes are related, which can happen when two communities are related or sub-communities of a larger group. For example, communities 4 and 6 in the *Reopen* dataset share many attributes. Lastly, there are differences across datasets. Communities in the *COVID* and *Reopen* datasets have strong prototypes which are generally isolated, there is a cluster of communities with inter-related prototypes in the *Election* dataset, and communities in the *Captain Marvel* dataset tend to have some common attributes with many other communities. We will now explore the underlying attributes which lead to these effects.

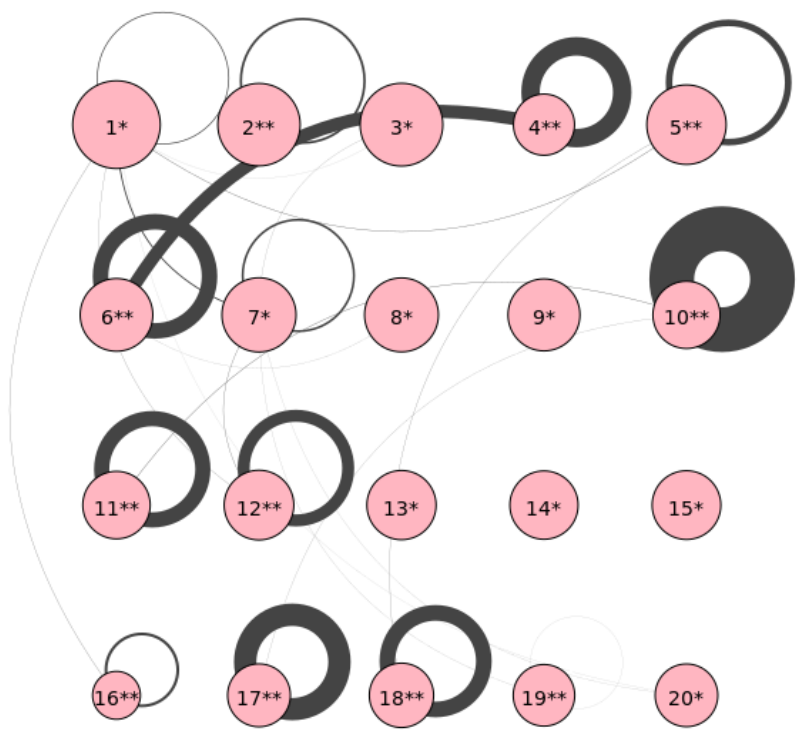

(a) Reopen

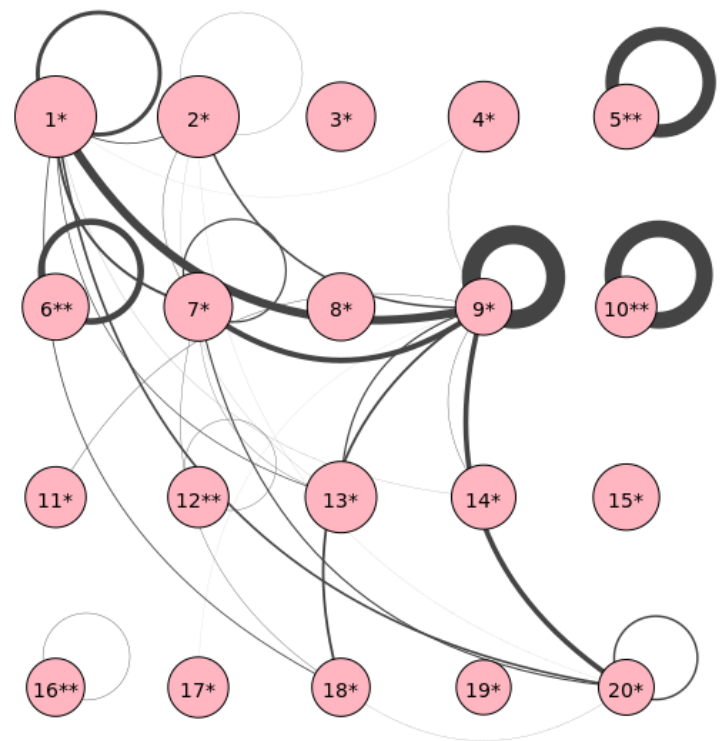

(b) Election

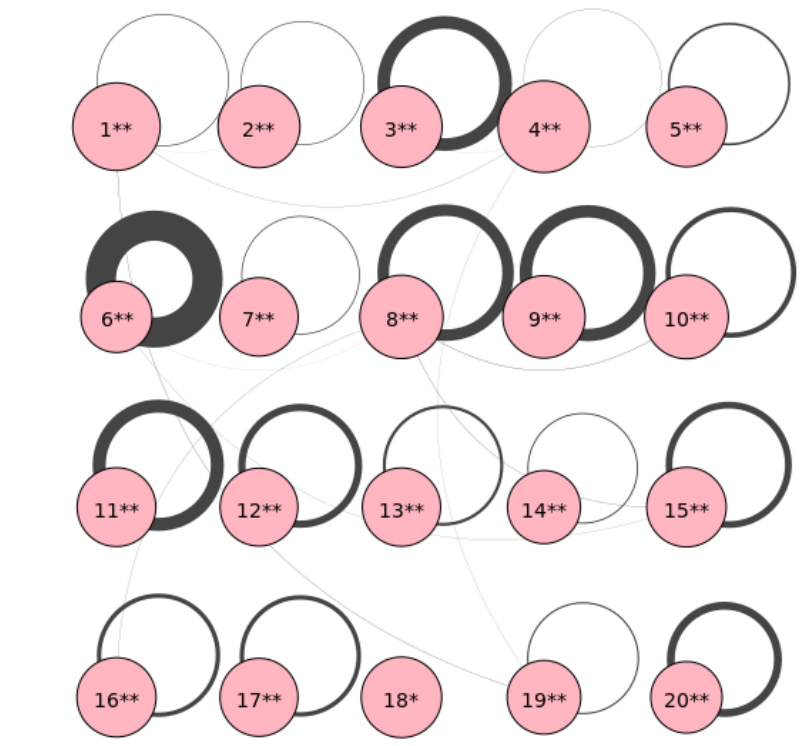

(c) COVID

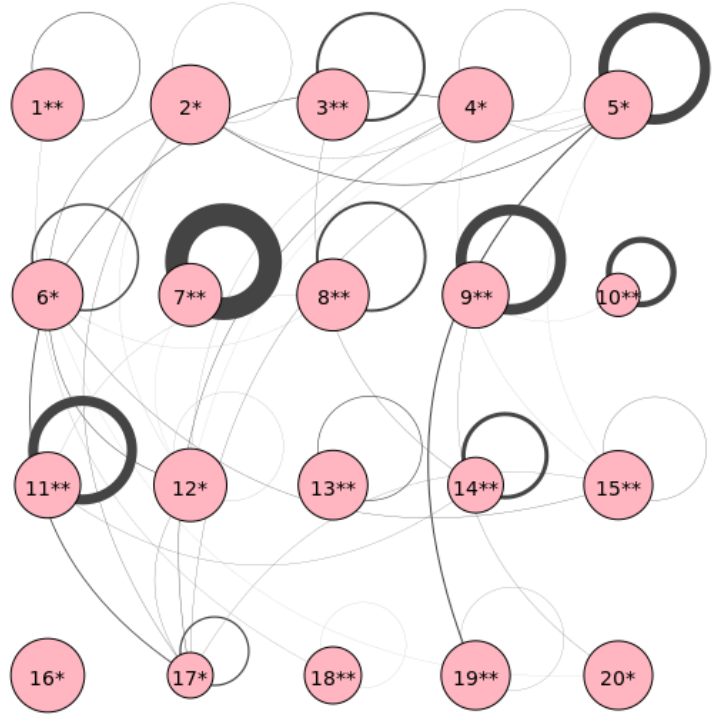

(d) Captain Marvel

**Fig. 3**: Following the process in Figure 1, observations for each dataset are shown. Each node is a community of users and the strength of connection between two nodes indicates the probability that users from those communities will share an attribute. If the probability of same-community members sharing attributes is higher than that of chance, the community is marked with a star. If the probability of a community member sharing attributes with non-community members is lower than that of chance, it is also marked with a star. If both are true, it is marked with two stars. Node size corresponds to the number of users in that community. For readability, a logarithmic scale is used, meaning that subtle node size differences correspond to drastic differences in community membership.

## The Construction of Prototypes

For a given community, $c$, the collection of attributes with highest values of $\overline{MV}_{c,j}$ in Equation 6 are taken to be the community's prototype. Again, the top 20 communities are analyzed for each dataset, ranked by their contribution to modularity, $V_{Q_c^P}$. The prototypes for the top four communities are displayed in Figure 4, and while those of the remaining communities and remaining datasets are displayed in Appendix B.3. Results are shown with *all attributes*, including those filtered in the previous analysis. While an attribute may not be salient in the overall network, it can still belong to a community's prototype.

We observe that prototypes are coherent representations of communities' multi-faceted identities which can be categorized into four dominant types: political, location or language, interest, and artificial. Political communities are those centered around specific politicians, political parties, or political ideologies. Examples of this include communities 1, 2, and 4 of the *Election* dataset, shown in Figure 4. Community 1 is a made up of Trump supporters, which predominantly differentiates itself with 🇺🇸 and #maga, but also shows support by mentioning General Flynn and Donald Trump directly. Community 2 is made up of Biden supporters which predominantly uses #resist and direct mentions of Joe Biden to differentiate itself, though it also uses hashtags like #bidenharris2020 and #blm. Its support of the Democratic Party is further shown with the use of 🌊, which indicates a "blue-wave," or a large Democratic turnout in the election. While the use of 🌊 is consistent with previous work documenting its usage in the American left, that work found that 🇺🇸 was a non-differentiator among pro and anti white-nationalist ideology [22]. However, previous work studying flag emojis specifically have found that the American flag is more popular among Republicans than Democrats [33]. Community 4 is made up of Black Lives Matter supporters, who often display she/her pronouns.

Using the *Election* prototypes in conjunction with the attribute-block diagram in Figure 3, it becomes clear that the cluster of related community prototypes (communities 1, 7, 9, 18, and 20), are all MAGA-related. While all are similar, each community tends to have a different focus. Community 7 is most defined by 🇺🇸, community 9 focuses on support of General Flynn, community 18 on the restart-leader account (a pro-MAGA Iranian political group), and community 20 on QAnon. The strength and prevalence of pro-MAGA community prototypes are greater than that of communities supporting Joe Biden or the Democratic party, especially in datasets that are less-directly political (*Captain Marvel* and *COVID*). This suggests that pro-MAGA users tend to be more isolated into conversational communities of similar users, even if they have multiple sub-communities.

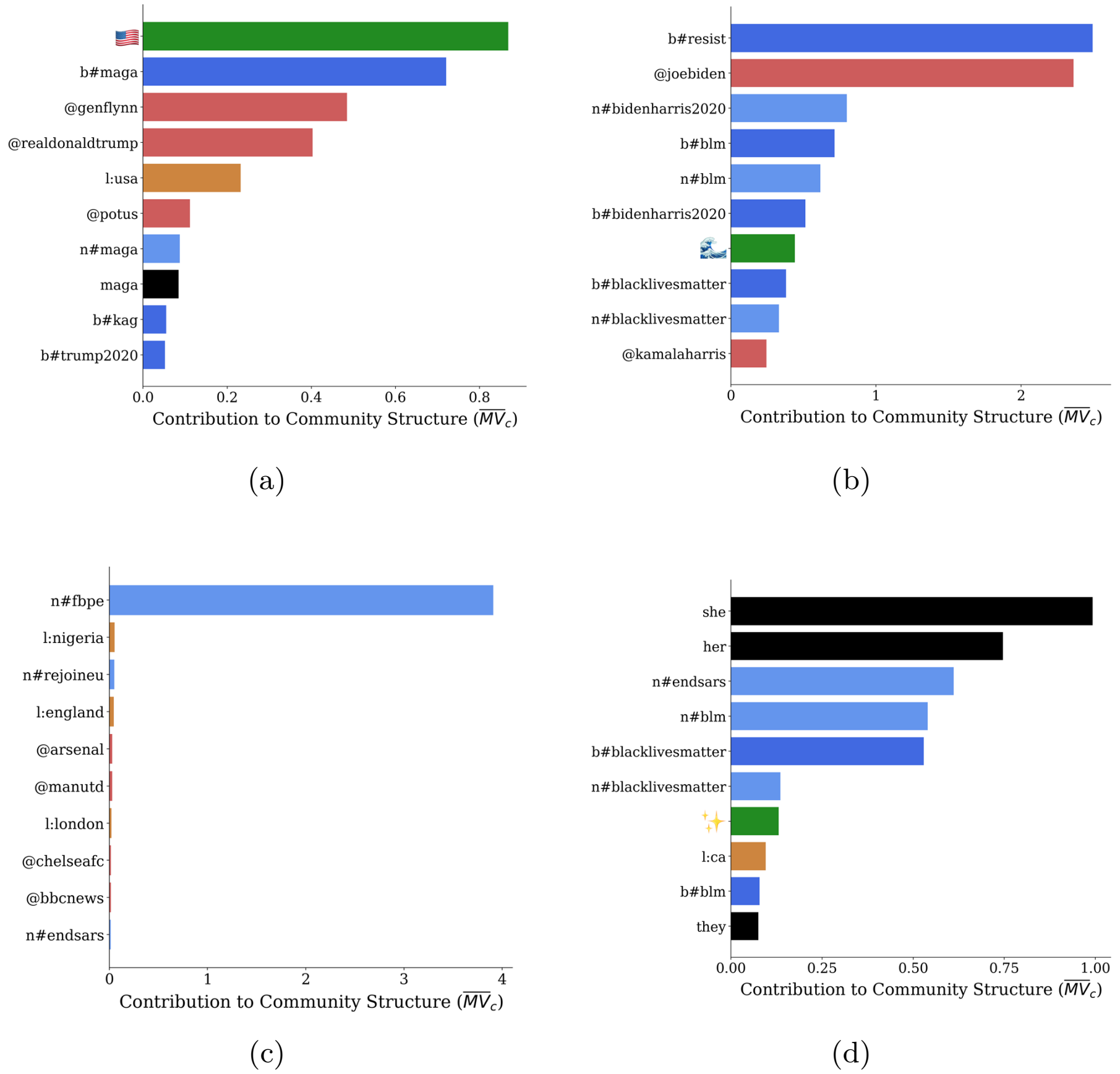

**Fig. 4**: Prototypes of the top four communities in the Election dataset. Colors emphasize different modalities. Prefixes are representative as follows: b#: bio hashtag, n#: name hashtag, l$: location unigram.

The second type of prototypes are those based on location and language. These prototypes are typically formed with country or city-based location unigrams, the flag of the country, and mentions of accounts which are related to the location. Two examples are communities 5 and 6 of the *Election* dataset, which are centered around India and France, respectively. In the case of Community 5, community members signal their Indian identity using "India" in their location, mentioning Prime Minister Modi's Twitter handle, and using 🇮🇳 in their biography. We observe that many of the communities with extremely strong prototypes, as visualized in Figure 3, are based on location or language. This could be because of the unified way of signaling identity (the names of countries and cities are agreed upon, unlike political hashtags), and because location-specific topics of discussion may be generating the communities. Communities in the *COVID* dataset are mostly of this type, explaining its strong and well-separated communities in Figure 3.

Third, many prototypes based on shared interests. The most common of such prototypes are K-pop fan groups, as seen in *Election* community 16, *Captain Marvel* communities 3 and 4, *COVID* communities 5 and 12, and *Reopen* community 17. Other interest-based groups include gamers, soccer fans, and TV show fans. The previously observed strong association between communities 4 and 6 in the *Reopen* dataset can now be understood. Community 4 is an interest-based community supporting the Indian actor, Vijay. Members of this community also tend to be Indian, creating attribute overlap with members of Community 6, a more general Indian-location-based community.

Lastly, there are "artificial" communities, wherein users signal their intention to create a community that can inflate the popularity and reach of its members. These may also be referred to as "follow-back" communities, since the users signal that if a member of the community follows them, they will reciprocate. The #fbpe (follow-back Pro-EU) community is present in all datasets (Community 3 in *Election*, Community 10 in *Captain Marvel*, Community 2 in *COVID*, and Community 5 in *Reopen*). Members of the community use hashtags in their username. This makes it easier for members to identifier each other, because name-hashtags can be seen without clicking into a users' biography. Generally, these communities do not have much in common beyond the community-signal itself. The recurrence of this community in all four datasets along with the size of the communities signal the prevalence of artificial communities in large Twitter discussions.

## Discussion

The main finding is that communication communities on Twitter do differentiate themselves via prototypes, as evidenced by the high levels of bipartite projected modularity in the multi-view attribute network. Further, we observe that these prototypes are multi-modal. That is, they are constructed using multiple types of attributes, including hashtags, mentions, and emojis in their biography, hashtags in their name, and unigrams in their location. It has been known that these types of attributes are used to signal users' social identity [11, 20, 25, 36, 46, 51, 53], but these findings indicate that this identity signaling is part of a larger group process which plays out within discourse communities.

This finding also strengthens the notion that automatically extracted clusters of users within Twitter communication networks can in fact be communities. While network clustering algorithms are often referred to as tools for community detection, a cluster of users which interact with themselves more than others is not necessarily a community in the psychological sense. As Turner has argued, shared self-definition through social attributes is more important for group membership than the structure of the group's interactions [58]. While clustering algorithms extracts groups who have interactional cohesion, they might not have shared interests, beliefs, or identities. In the datasets we examined, members of communication clusters *do* signal beliefs, interests, and identities which help form a stronger basis of community. Recent work has suggested that the follower network can be partitioned into interest-based groups or "flocks" which can be used to understand public opinion [64]. Our findings suggest that this may also be done using communication clusters, which are more dynamic and can be collected on specific discussions.

Although the clusters within datasets that we examined have cohesive beliefs, interests and identities, it is likely that there are exceptions. With the methods and code that we develop in this work, the cases when this does and does not occur can be distinguished. Studies beyond the scope of political discussion are called for to understand the factors which affect the exhibition or strength of community prototypes.

Our analysis also shows that the prototypes of individual communities shed light on their membership's identities and beliefs. Because we are studying political datasets, it is natural that the prototypes are political in nature. However, the strength of prototypes and their starkly opposed political affiliations point at a form of political polarization not typically measured. Polarization is often studied using either the stance of users on specific issues [8, 21, 42], or the content that they retweet [15]. Here, we see polarization in terms of identity: the presence of political community prototypes indicates that the discussions between users who identity as MAGA Republicans and those who identify as the Democratic Resistance are largely separated. Future work using this framework to study interactions between users across polarized communities is of interest. More generally, using this framework to quantify a community member's alignment with their prototype is of interest, given this alignment's crucial role within the social identity perspective [54]. Studies in this direction could add granularity to the recent finding that identity cues have significant effects on users' comment voting behavior on a social media site similar to Reddit [55]. Identity cues encoded in Twitter community prototypes are much stronger than those seen on sites like Reddit, and the ability to measure identity alignment could distinguish between different types of effects.

An important limitation is the inability to attribute the causal mechanism behind these community prototypes. Users may signal their identity attributes to gain popularity or acceptance within a community. The effect of the follower network on these outcomes is also of interest for future work, since it naturally biases the interaction network. It has recently been shown that users who follow political elites on Twitter overwhelmingly follow those from only their ideological in-group [60]. Further, users may also choose to follow or unfollow users based on their displayed attributes; it has been shown that users may choose to unfollow, block, or mute users outside of their ideological in-group during times of polarization [7].

Beyond following networks, the interaction between user profiles and Twitter recommendations is an important factor which is challenging to study due to the proprietary nature of the platform's design. It is possible that Twitter's following recommendation algorithm leverages profile attributes to recommend that users with similar profiles follow each other. Such an algorithm could strengthen community prototypes. Further, the utilization of user profile similarity in content recommendation could encourage users with similar profiles to engage with each other, which would also strengthen community prototypes. The deployment of such algorithms could have large impacts on the structure and dynamics of online discourse communities. This includes the possibility of increasing levels of polarization in political discourse. This social process with algorithmic feedback could give a more specific mechanism driving the recently-named partisan sorting phenomena, where previously separate social divisions have become aligned on the basis of individual's social identity [56]. Investigations into the usage of such recommendation and their interplay with the group processes governing the creation and adoption of community prototypes is of interest for future work.

## Conclusions

The social identity perspective has been a successful approach to studying problems relating to social media communities like polarization and radicalization [1, 23, 27, 31]. This perspective relies heavily on the concept of community prototypes, or collections of attributes differentiating a community. However, it has previously been unclear if online conversational communities exhibit prototypes due a lack of appropriate methodology. We develop such a methodology by modeling user-attribute relationships with a multi-view bipartite projection network, applying bipartite projection modularity, and developing bipartite projection modularity vitality.

Applying the approach to four Twitter datasets ranging from roughly 4 to 30 million tweets, we demonstrate the presence of multi-modal community prototypes within US-focused political discussions. We find that communities strongly differentiate themselves through hashtags, mentions, and personal identifiers, and emojis that their members put in their biographies, as well as unigrams they put in their location field. Examining the makeup of these prototypes, we find there are 4 major types of communities: political, location/language-based, interest-based, and artificial. Multiple community prototypes are seen to re-occur across datasets, including MAGA Republicans, Resist Democrats, and K-pop fans.

The presence of community prototypes points to potential problems and to research opportunities. While the construction of prototypes may be a natural sociological phenomenon, the high levels of contrast between ideologically opposed communities may indicate identity-related polarization. Further, the potential presence of such polarization realized through features which may be leveraged through highly sophisticated but hard to interpret recommendation engines is cause for concern which warrants future investigation. Outside of polarization, the methods developed for uncovering community prototypes enables a large body of future work testing social theory regarding how the alignment between individuals and their community prototypes relates to the nature of their interactions. By operationalizing the detection of prototypes in online communities, we hope this work can pave the way in translating a wide range of social identity results to the online setting.

## Acknowledgements

This work was supported in part by the Office of Naval Research (ONR) MURI: Persuasion, Identity, & Morality in Social-Cyber Environments Award N00014-21-12749, Scalable Tools for Social Media Assessment Award N00014-21-1-2229, and Group Polarization in Social Media Award N00014-18-12106. It was also supported by the Center for Computational Analysis of Social and Organization Systems (CASOS). Thomas Magelinski was also supported by the IDeaS Center as a Knight Fellow. The views and conclusions contained in this document are those of the authors and should not be interpreted as representing the official policies, either expressed or implied, of the ONR.

# Appendix A Datasets

Each dataset was collecting using a keyword search within Twitter's API[3].

The *Reopen America* dataset was originally collected by Babcock and Carley, and aims to capture discussion around the Reopen America protests occurring in the summer of 2020 [3]. These were a series of decentralized protests across the United States in the summer of 2020, where protesters sought to "reopen the economy" by removing COVID-related restrictions. The largest of these protests occurred in Michigan, and was named "operationgridlock," due to the protestors tactics of parking cars to create gridlock traffic. President Donald Trump tweeted in support of these protests using the term "liberate," making it a rallying cry for supporters. Because of this, the dataset was collected using "openup", "reopen", "operationgridlock", and "liberate". Additionally, all of the US State abbreviations were appended to each of the search terms, e.g., "liberateNY", to collect the Reopen America protest for the corresponding state. The search was performed from April 1, 2020, to June 22, 2020.

The *Election* dataset was collected to capture discussion of the 2020 US presidential election. The dataset was collected using a keyword search on election-related terms as well as the handles of prominent politicians. The collection began the day before the election, November 2, 2020, and continued through the day after most major news outlets had called the election in Biden's favor, November 8, 2020. The list of keywords used is as follows: #election2020, #presidentialelection, #democrats, #republicans, #JoeBiden, #BidenHarris2020, #Biden, #MAGA, #KAG, #VoteByMai, #USPS, #SaveTheUSPS, #voterfraud, #BlackLivesMatter, #BLM, #reopen, #reopenamerica, #IranSanctions, #QAnon, #WWG1WGA, natural born, @JoeBiden, @realDonaldTrump, @POTUS, @Mike_Pence, @VP, @KamalaHarris, @SenKamalaHarris, @USPS, @CoryGardner, @SenCoryGardner, @Hickenlooper, @Perduesenate, @sendavidperdue, @ossoff, @joniernst, @SenJoniErnst, @GreenfieldIowa, @SenSusanCollins, @SenatorCollins, @SaraGideon, @SteveDaines, @stevebullockmt, @GovernorBullock, @ThomTillis, @SenThomTillis, @CalforNC.

The *COVID* dataset was collected to capture the general discourse around the coronavirus right after it was officially declared a pandemic. Thus, dataset collection began on March 11, 2020, the day WHO declared a pandemic, and continued through March 17, 2020. The list of keywords used is as follows: coronaravirus, coronavirus, wuhan virus, wuhanvirus, 2019nCoV, NCoV, NCoV2019, covid-19, covid19, covid 19.

The *Captain Marvel* dataset was originally collected by Babcock and Carley, and aims to capture discussion around the premier of Captain Marvel, Marvel's first female-led superhero movie [4]. The main keywords used for dataset collection include: #CaptainMarvel, Captain Marvel, Brie Larson, Alita, SJW, Feminazi, #BoycottCaptainMarvel, and #AlitaChallenge. The initial goal of this dataset collection was to study the development of misinformation and counter-narratives around the movie. Because of this, the collection procedure is more complex than that of the other datasets, and we refer to the original paper for those details.

[3] https://developer.twitter.com/en/docs/twitter-api

### Network Construction and Communication Communities

Communication communities were derived as follows. First, a communication network between users was constructed. This network recorded each interaction between users, with the following actions counting as interactions: reply, mention, retweet, and quote. Combinations of actions were also considered. For example, if a user retweets a reply, that user is connected to both the original author, and the user that was being replied to. These interactions were combined into an undirected user-to-user network, where edge weights indicate the number of interactions between a pair of users. Network statistics for each dataset can be found in Table 1. Finally, the Leiden algorithm maximizing modularity was used to uncover communication communities [57]. We note that while the practice of modularity maximization is an extremely popular method it has been criticized from the point of view of inferential network analysis due to its inability to distinguish statistically significant communities from noise, and due to its glassy nature as an objective function [19, 35, 47]. The size of the communication networks prohibits the use of some powerful inferential techniques developed to tackle these problems. Because we are not aiming to make statistical claims about the structure of the online communities, only about their prototypes, we continue with the Leiden approach.

# Appendix B Extended Results Diagrams and Tables

Here we display the tables and Figures that could not be fit in the main article body.

## B.1 Community Diagram on Unfiltered Data

## B.2 Salient Attributes

Tables of the most and least salient *biography* attributes for the *Reopen*, *COVID*, and *Captain Marvel* datasets are given in Tables B1, B3, and B5, respectively. Accordingly, the tables for *non-biography* attributes are given in Tables B2, B4, and B6.

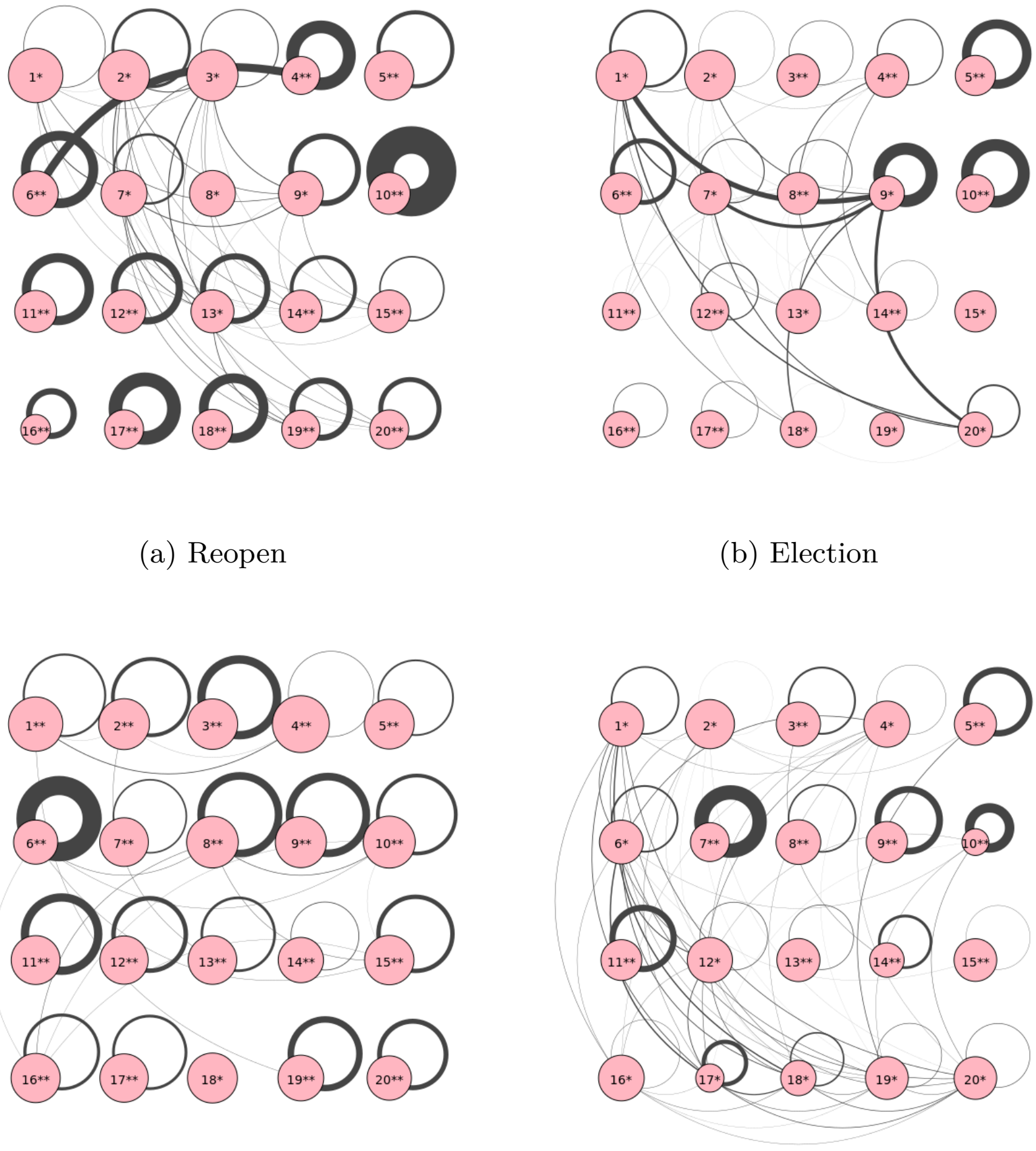

(a) Reopen (b) Election

(c) COVID (d) Captain Marvel

**Fig. B1**: The community-to-community shared-attribute relationships are shown just as in Figure 3, using the *unfiltered* data.

| Personal ID | | Mention | | Hashtag | | Emoji | |
|---|---|---|---|---|---|---|---|
| **S** | **NS** | **S** | **NS** | **S** | **NS** | **S** | **NS** |
| she | writer | @genflynn | @manutd | #maga | #blacklivesmatter | 🇺🇸 | ❤️ |
| her | he | @actorvijay | @nytimes | #kag | #blm | ✨ | 💙 |
| maga | him | @realdonaldtrump | @lfc | #wwg1wga | #resist | 🌊 | 🤍 |
| they | husband | @potus | @arsenal | #trump2020 | #resistance | 🇮🇳 | 💕 |
| black lives matter | wife | @salesforce | @chelseafc | #followbackhongkong | #fbr | ✝️ | 🌈 |

**Table B1**: The most salient (S) and least salient (NS) attributes of each attribute derived from user biographies within the *Reopen* Dataset

| **Name Hashtag** | | **Location Unigram** | |
|---|---|---|---|
| **S** | **NS** | **S** | **NS** |
| #blm | #blacklivesmatter | england | usa |
| #fbpe | #acab | india | new |
| #maga | #stayhome | london | united |
| #wwg1wga | #bim | uk | ca |
| #kag | #junkterrorbill | africa | the |

**Table B2**: The most salient (S) and least salient (NS) attributes of each attribute *not* derived from user biographies within the *Reopen* Dataset

| **Personal ID** | | **Mention** | | **Hashtag** | | **Emoji** | |
|---|---|---|---|---|---|---|---|
| **S** | **NS** | **S** | **NS** | **S** | **NS** | **S** | **NS** |
| she | ig | @flamengo | @bts_twt | #maga | #blacklivesmatter | 🇺🇸 | ❤️ |
| her | instagram | @vascodagama | @manutd | #kag | #mufc | ✨ | 🤍 |
| they | writer | @fluminensefc | @lfc | #resist | #ynwa | 🇫🇷 | 💙 |
| maga | music | @realdonaldtrump | @realmadrid | #trump2020 | #bernie2020 | 🇮🇳 | ⚽ |
| he | fan account | @narendramodi | @fcbarcelona | #wwg1wga | #bts | 🇲🇽 | 💛 |

**Table B3**: The most salient (S) and least salient (NS) attributes of each attribute derived from user biographies within the *COVID* Dataset

| **Name Hashtag** | | **Location Unigram** | |
|---|---|---|---|
| **S** | **NS** | **S** | **NS** |
| #oustduterte | #loona1stwin | argentina | usa |
| #yoapruebo | #fbpe | france | de |
| #apruebo | # | brasil | new |
| #facciamorete | #bernie2020 | españa | the |
| #maga | #flattenthecurve | india | ca |

**Table B4**: The most salient (S) and least salient (NS) attributes of each attribute *not* derived from user biographies within the *COVID* Dataset

| **Personal ID** | | **Mention** | | **Hashtag** | | **Emoji** | |
|---|---|---|---|---|---|---|---|
| **S** | **NS** | **S** | **NS** | **S** | **NS** | **S** | **NS** |
| she | gamer | @weareoneexo | @bts_twt | #maga | #resist | 🇺🇸 | ❤️ |
| her | writer | @actorvijay | @twitch | #kag | #blacklivesmatter | 🏳️‍🌈 | 🤍 |
| fan account | music | @genflynn | @manutd | #2a | #marvel | 😎 | 💙 |
| fub free | ig | @b_hundred_hyun | @marvel | #trump2020 | #blm | 🐰 | ✨ |
| maga | artist | @iamsrk | @lfc | #nra | #mufc | 🇲🇾 | 💕 |

**Table B5**: The most salient (S) and least salient (NS) attributes of each attribute derived from user biographies within the *Captain Marvel* Dataset

| **Name Hashtag** | | **Location Unigram** | |
|---|---|---|---|
| **S** | **NS** | **S** | **NS** |
| #saveodaat | #twoofus | malaysia | usa |
| #releasethesnydercut | #fightforwynonna | brasil | the |
| #maga | #renewodaat | france | ca |
| #fbpe | #saveshadowhunters | thailand | new |
| #peoplesvote | #savedaredevil | indonesia | england |

**Table B6**: The most salient (S) and least salient (NS) attributes of each attribute *not* derived from user biographies within the *Captain Marvel* Dataset

## B.3 Prototypical Attributes

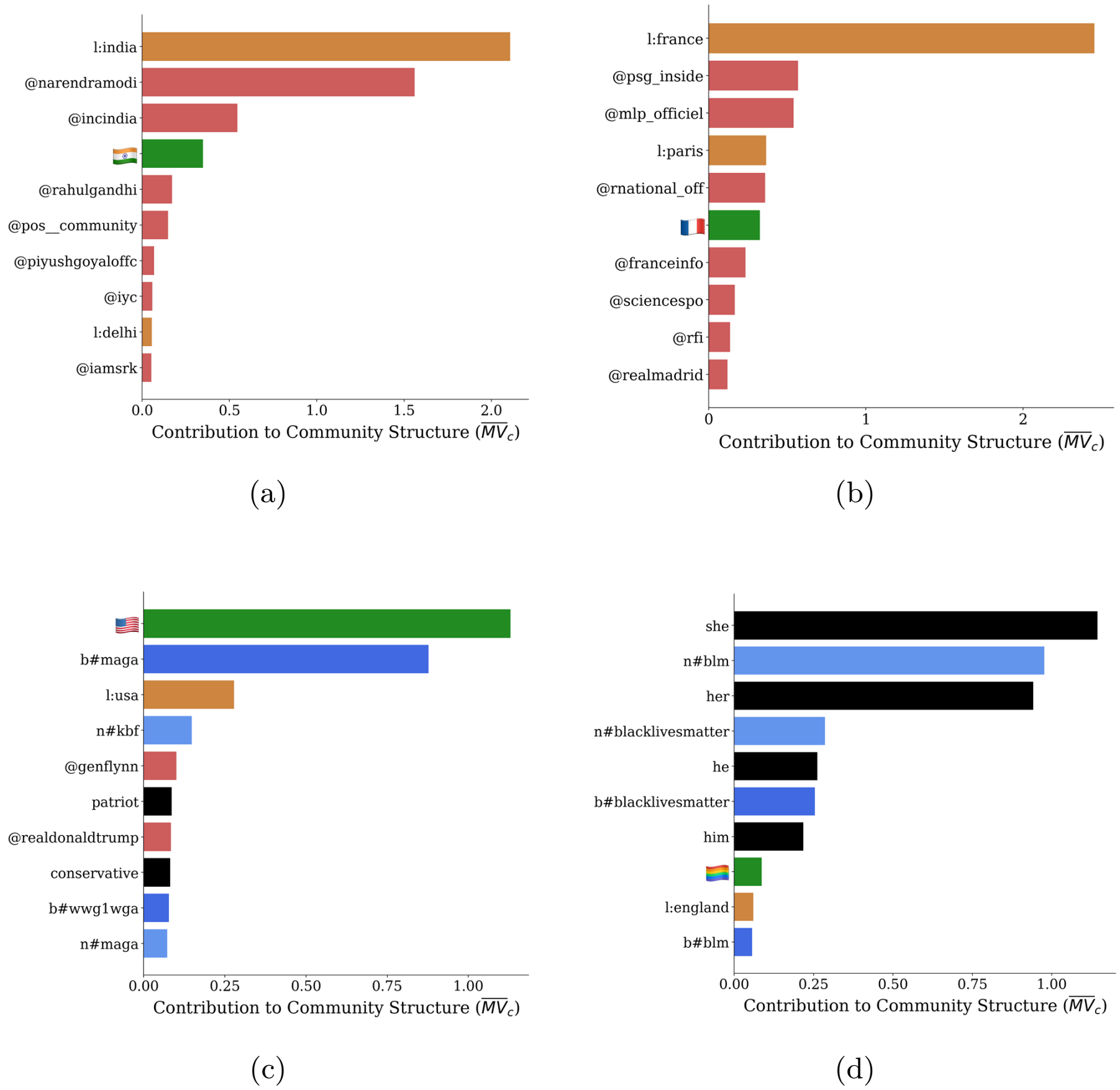

**Fig. B2**: Prototypes of the communities 5-8 in the *Election* dataset.

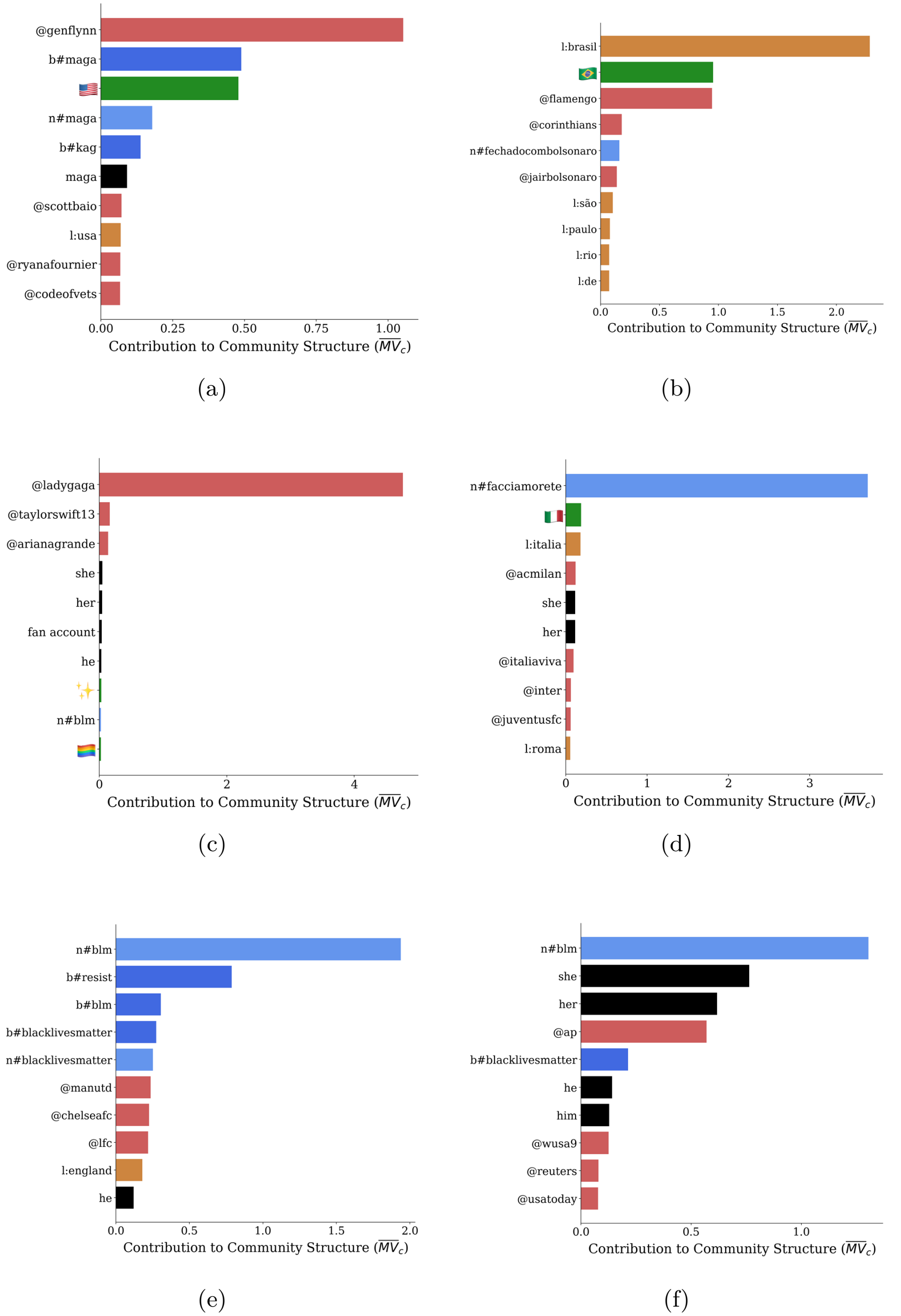

**Fig. B3**: Prototypes of the communities 9-14 in the *Election* dataset.

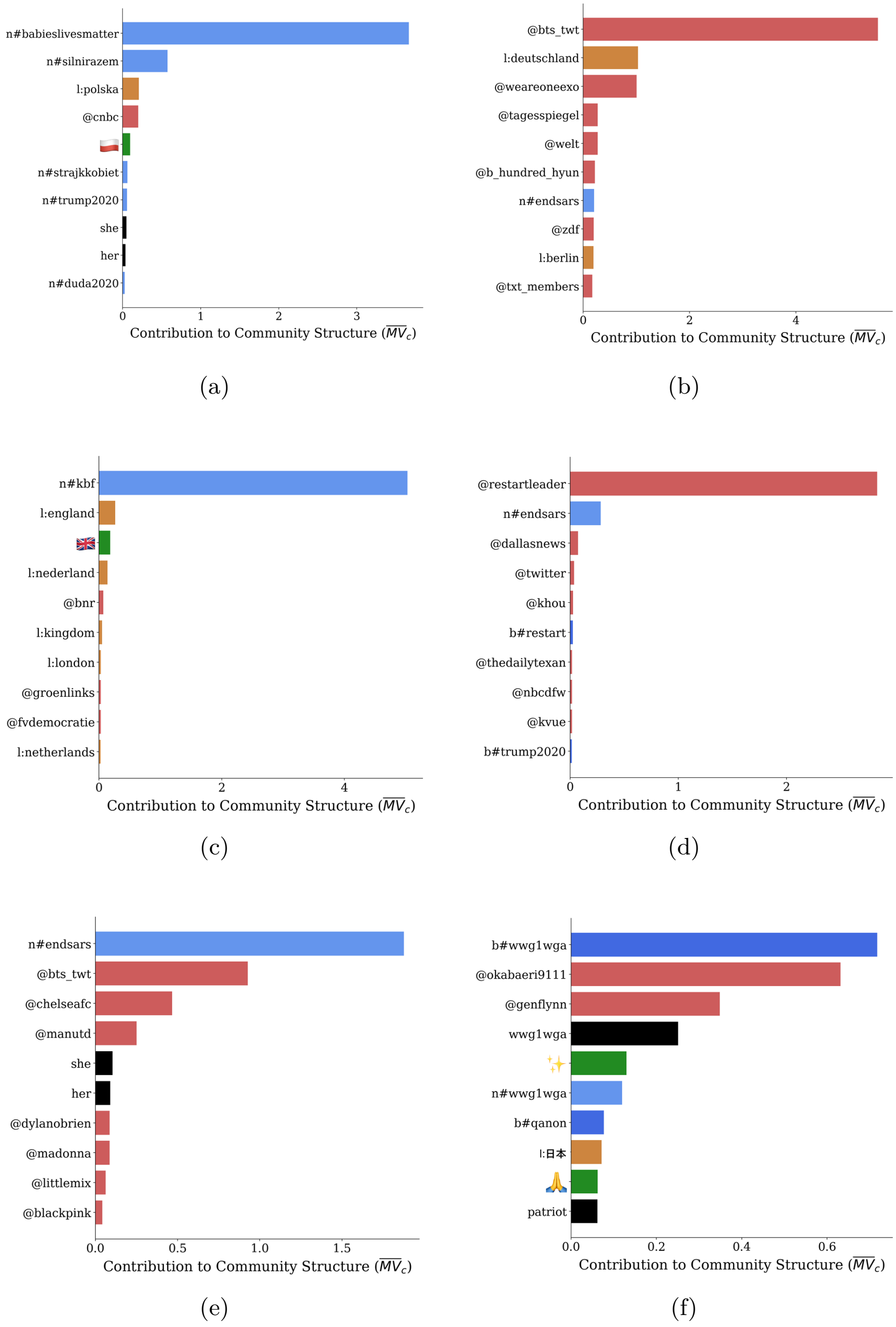

**Fig. B4**: Prototypes of the communities 15-20 in the *Election* dataset.

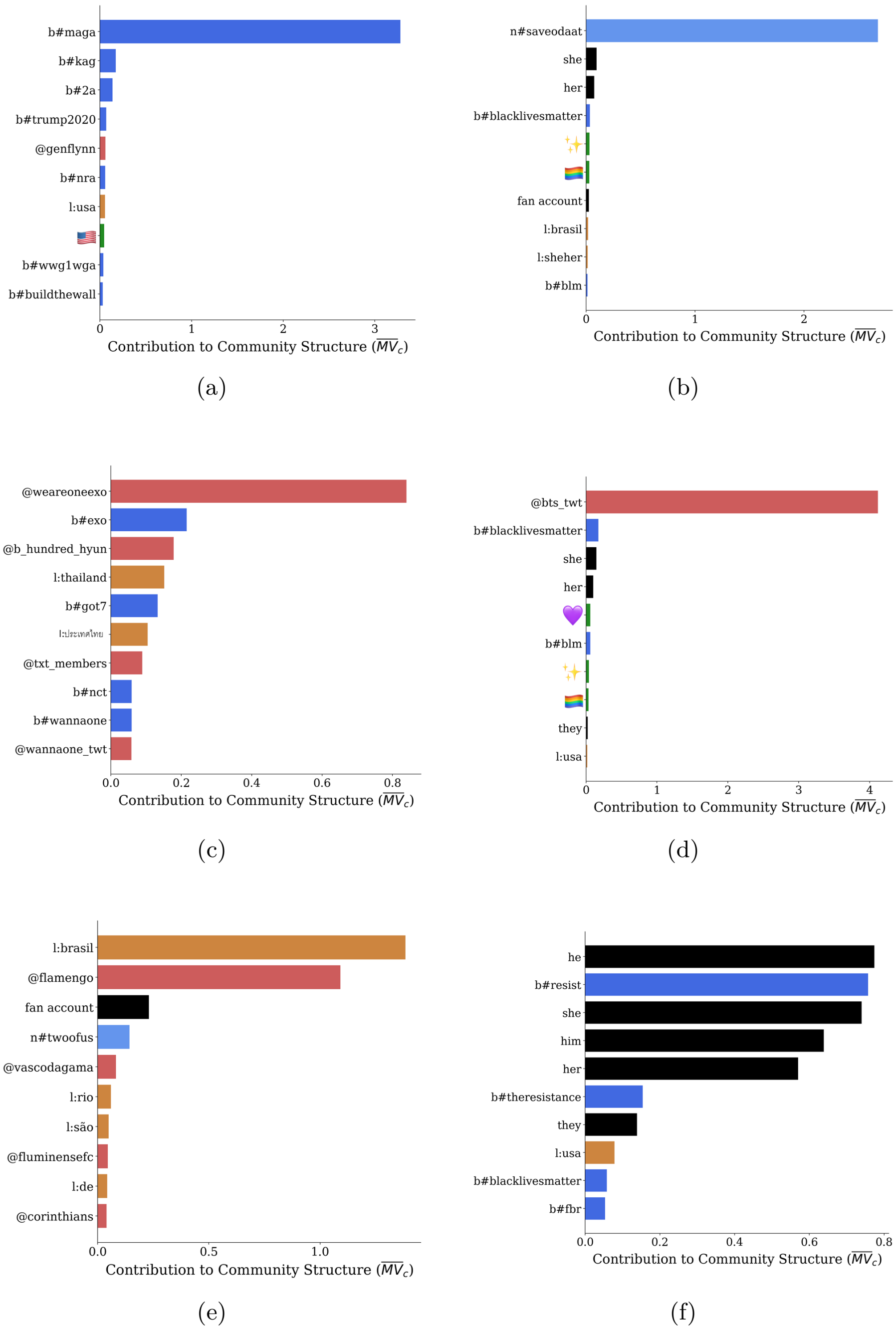

**Fig. B5**: Prototypes of the communities 1-6 in the *Captain Marvel* dataset.

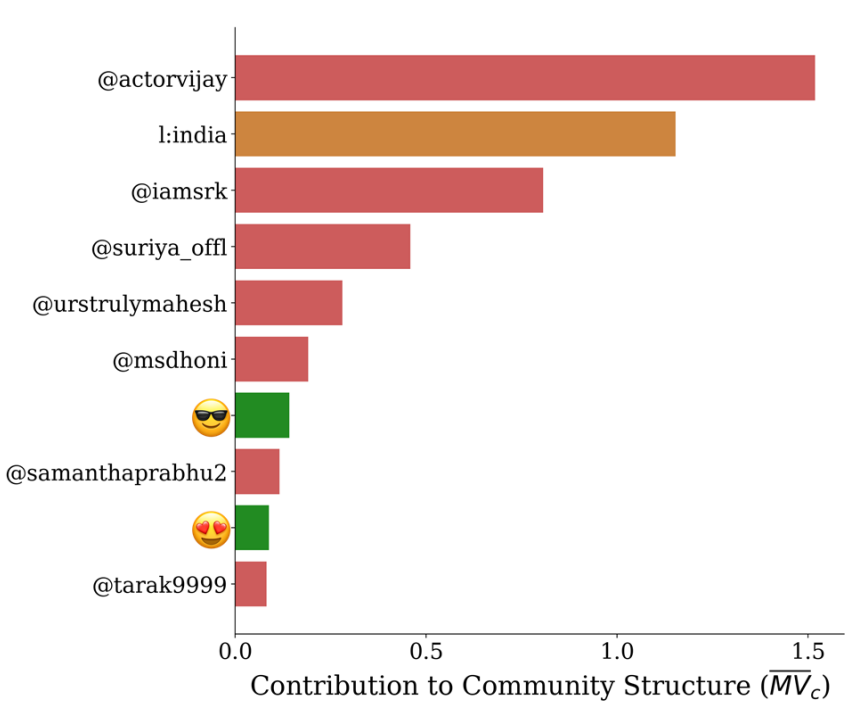

(a)

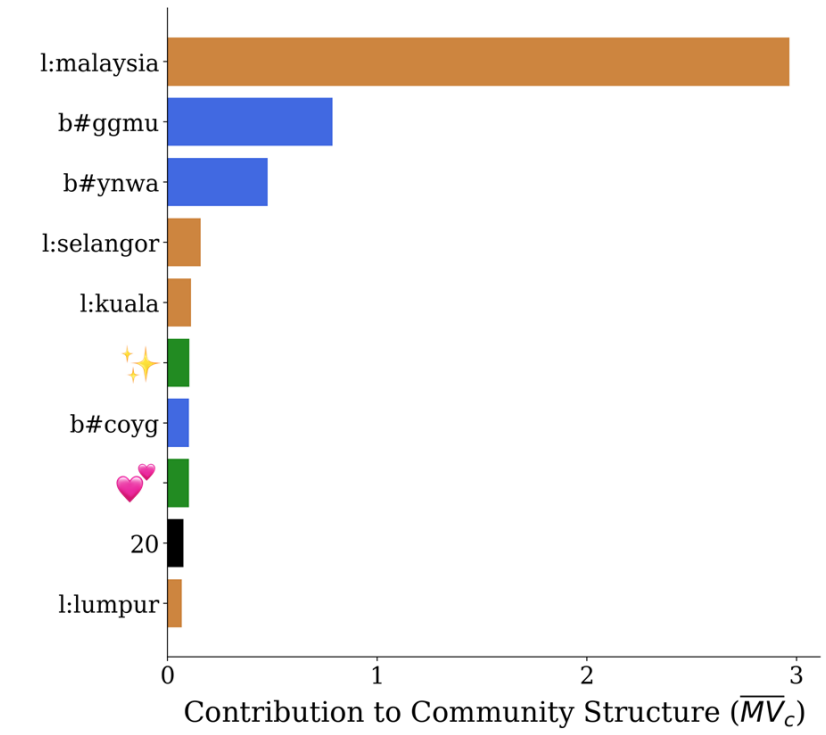

(b)

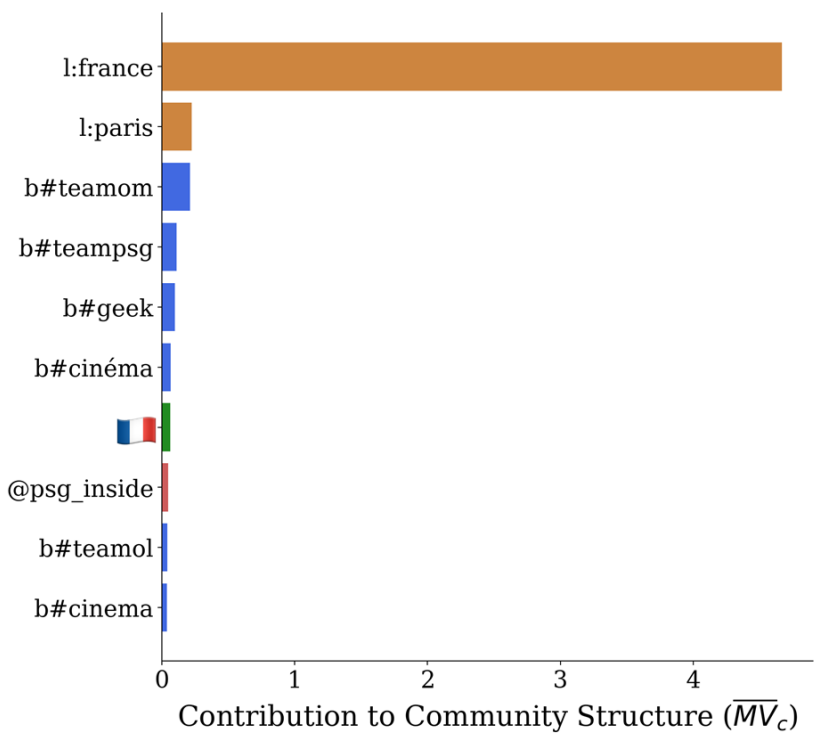

(c)

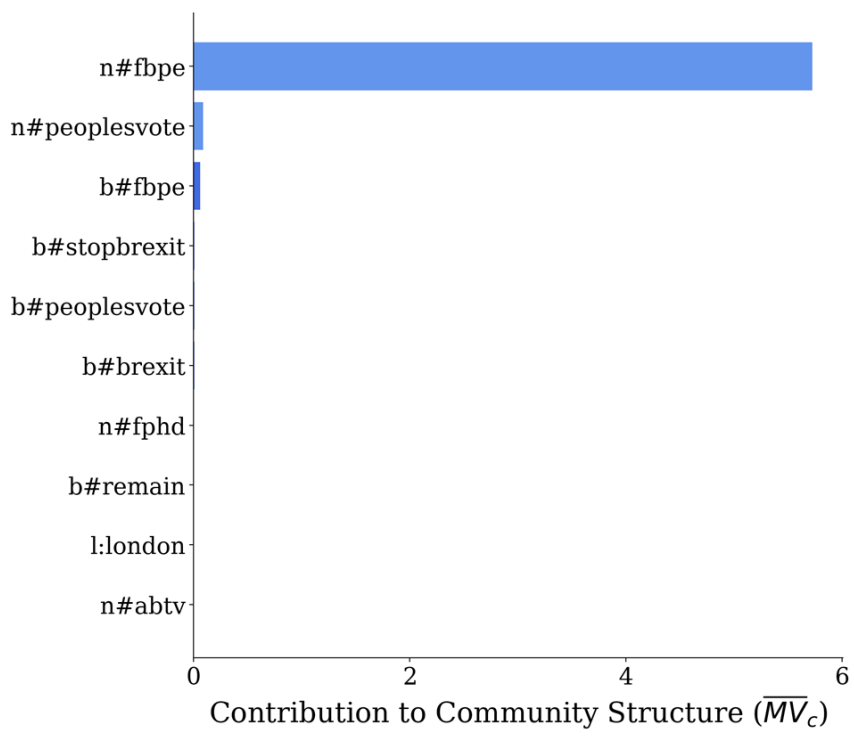

(d)

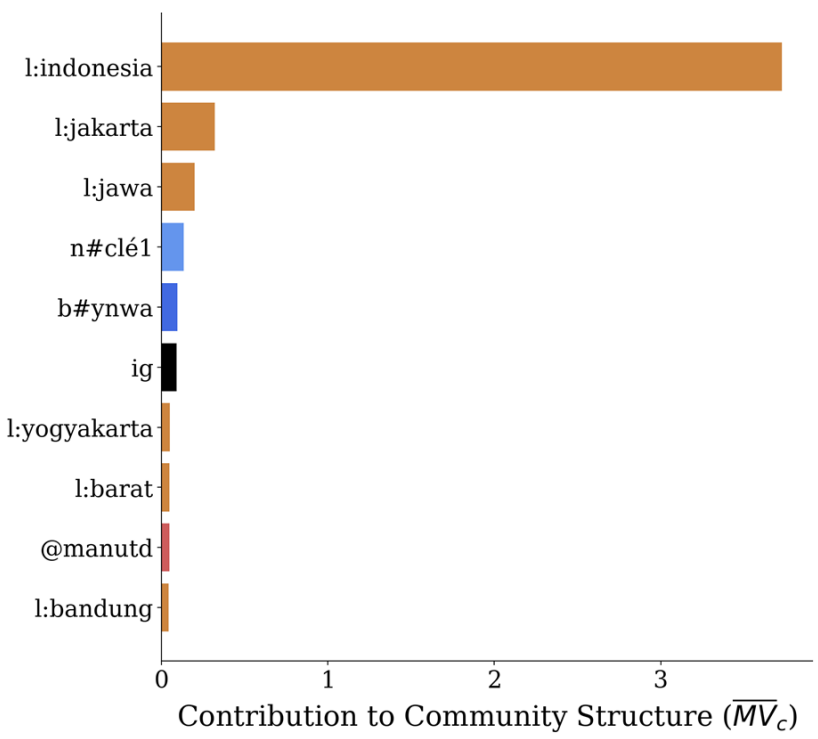

(e)

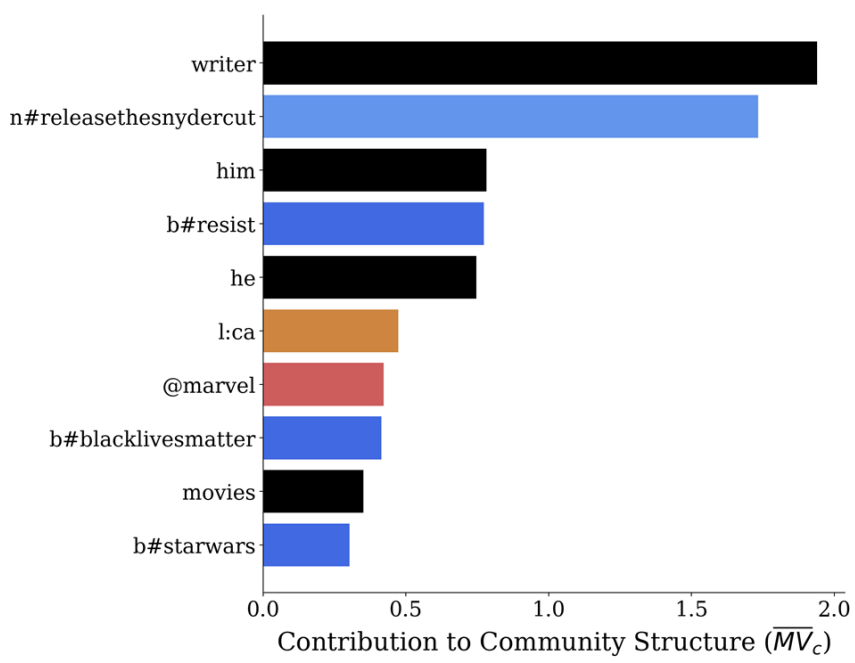

(f)

**Fig. B6**: Prototypes of the communities 7-12 in the *Captain Marvel* dataset.

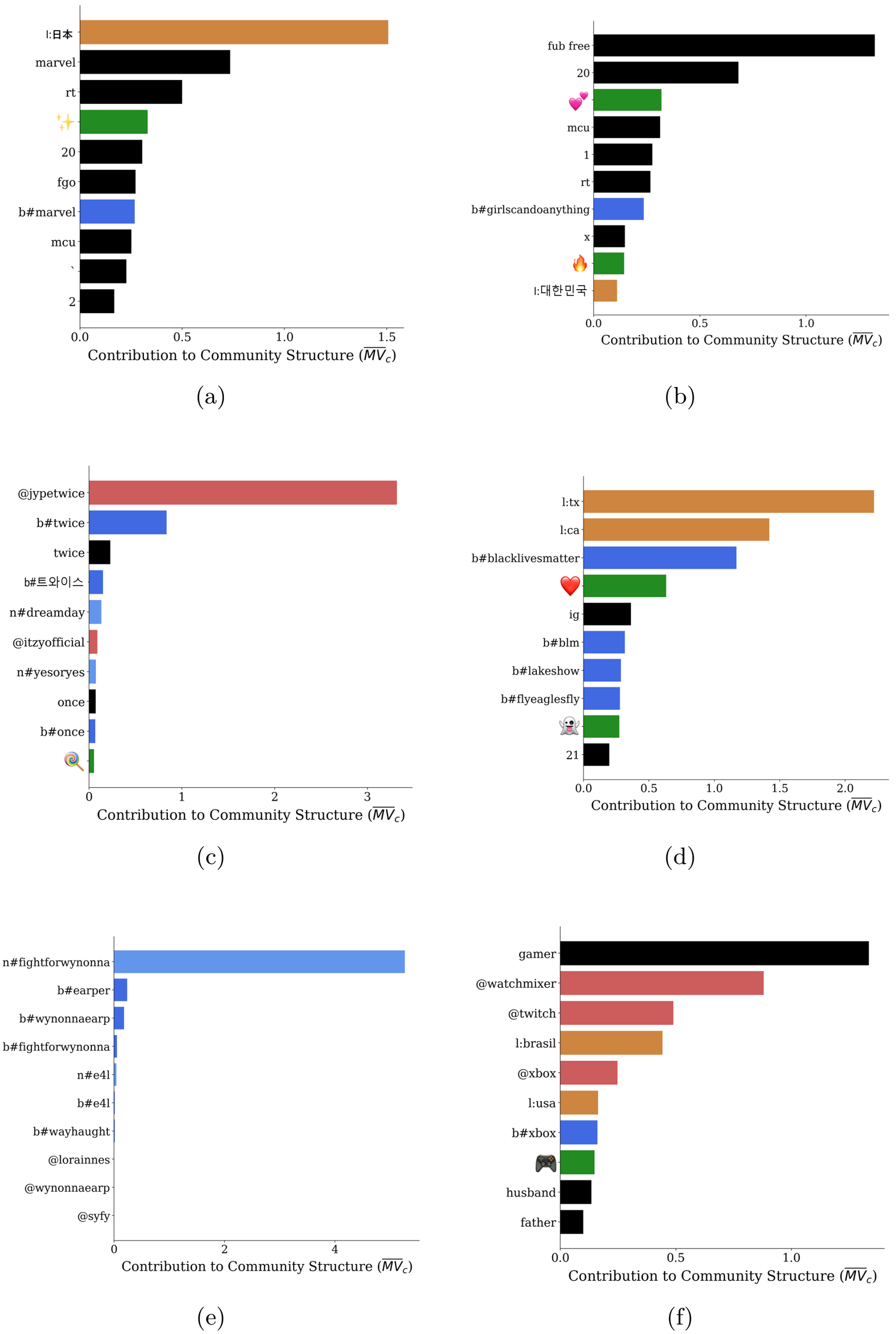

**Fig. B7**: Prototypes of the communities 13-18 in the *Captain Marvel* dataset.

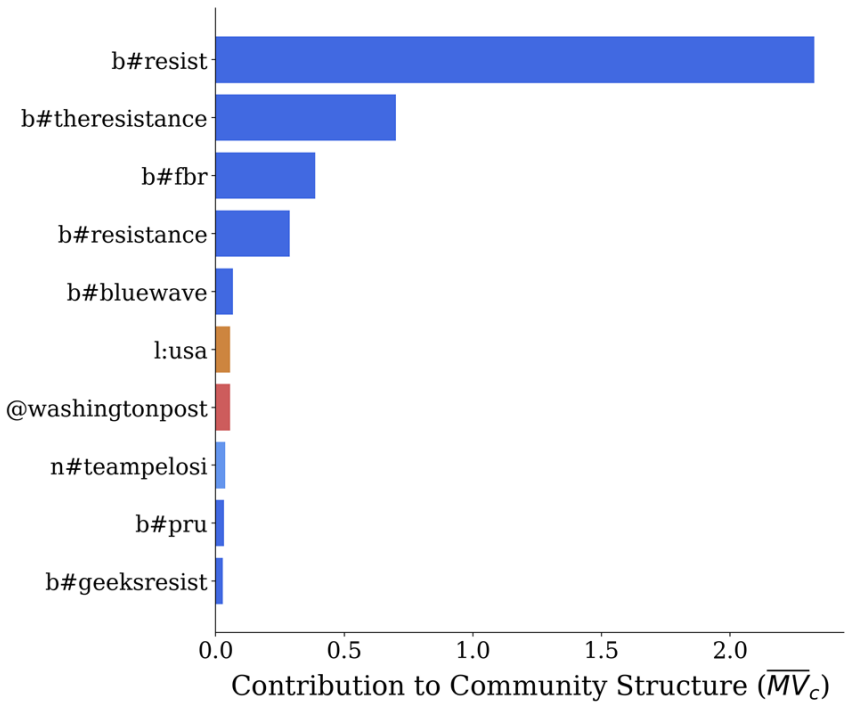

(a)

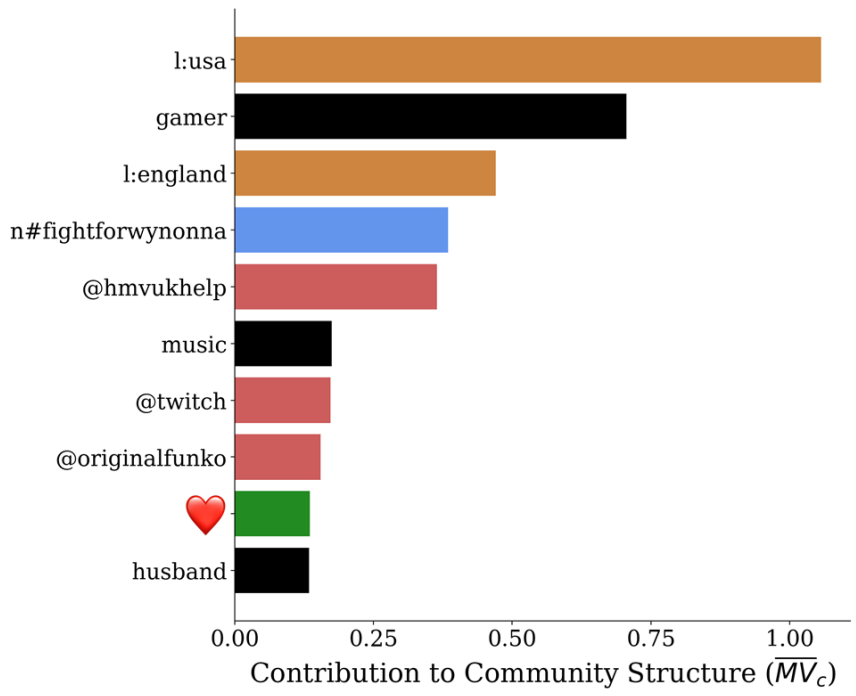

(b)

**Fig. B8**: Prototypes of the communities 19-20 in the *Captain Marvel* dataset.

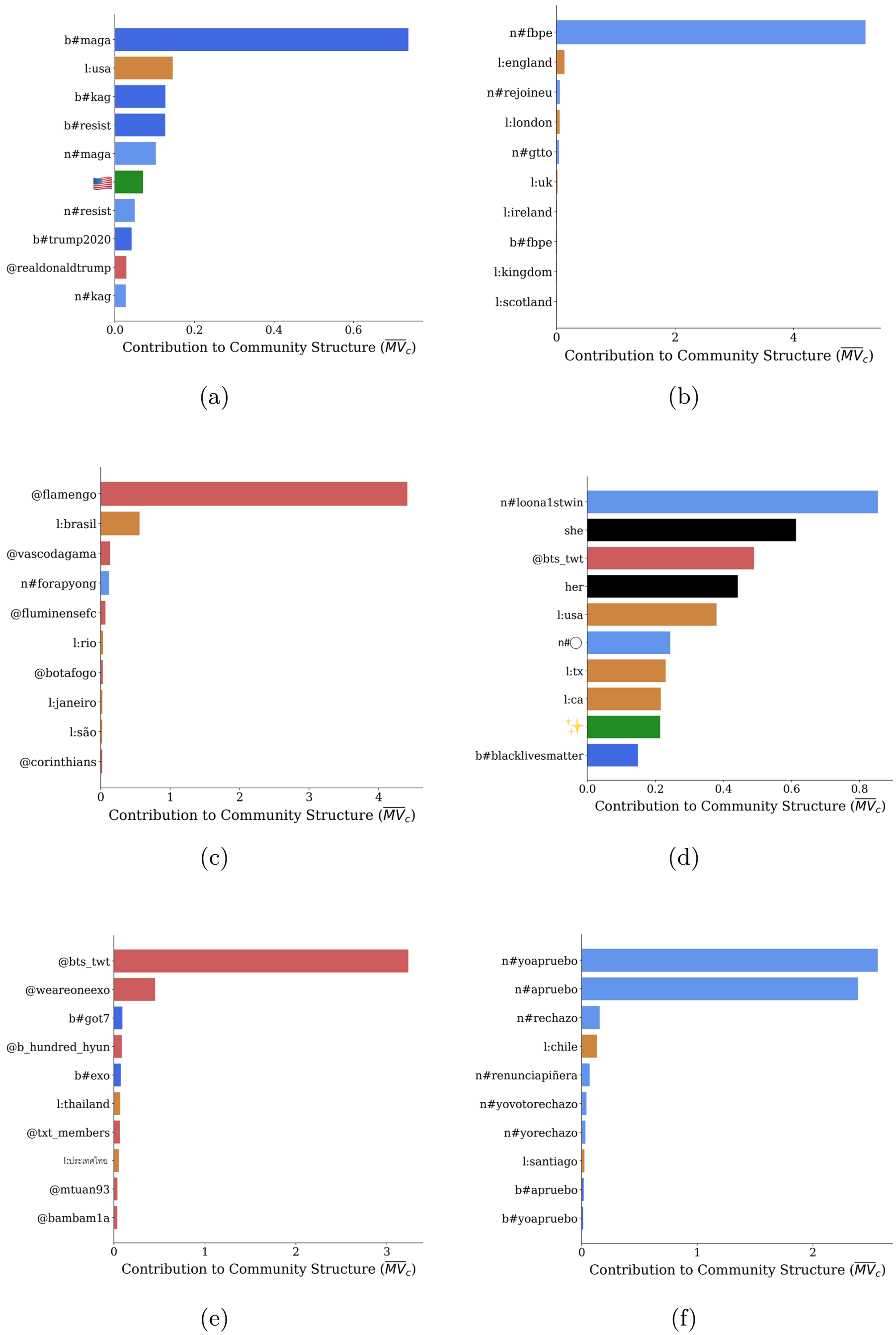

**Fig. B9**: Prototypes of the communities 1-6 in the *COVID* dataset.

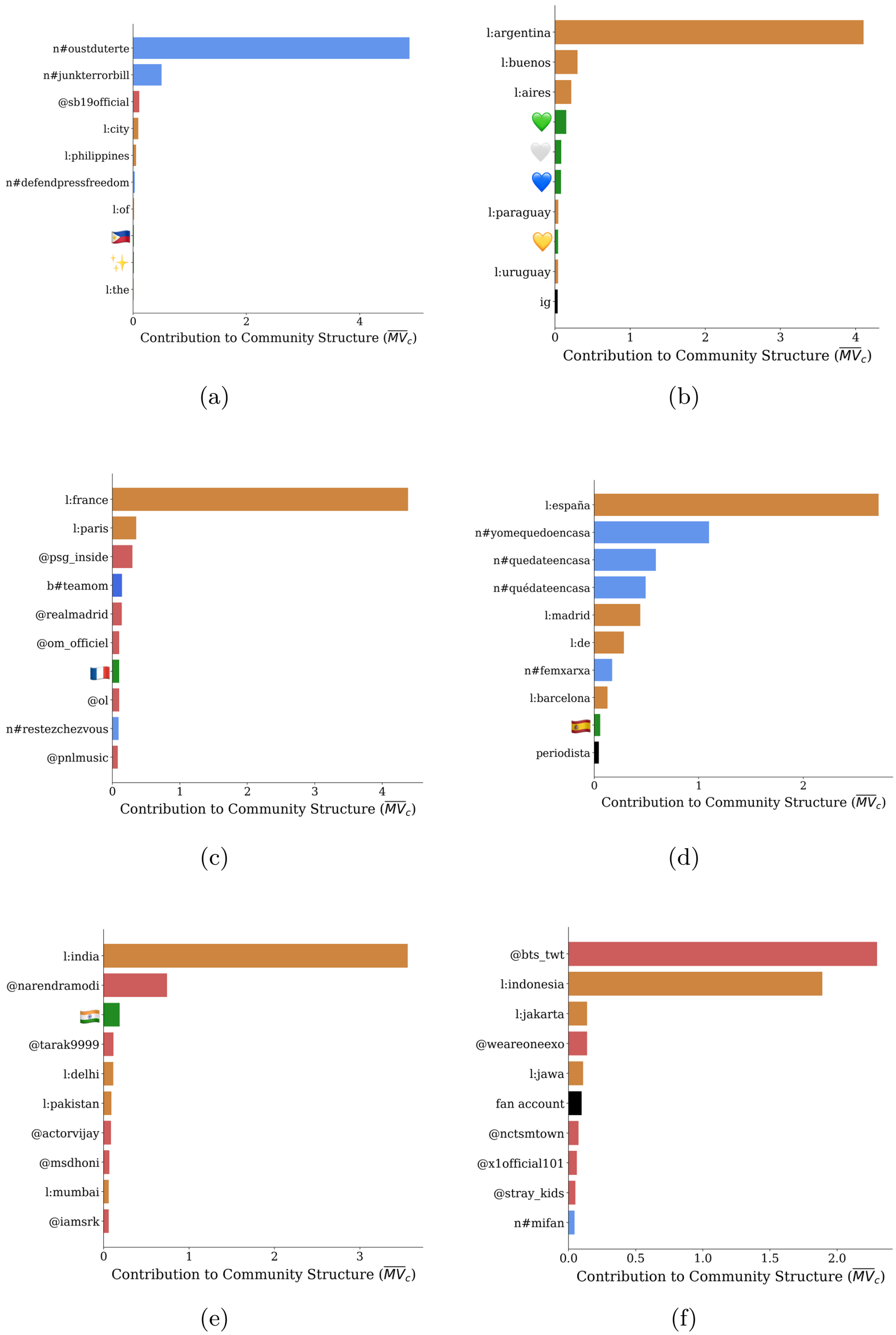

**Fig. B10**: Prototypes of the communities 7-12 in the *COVID* dataset.

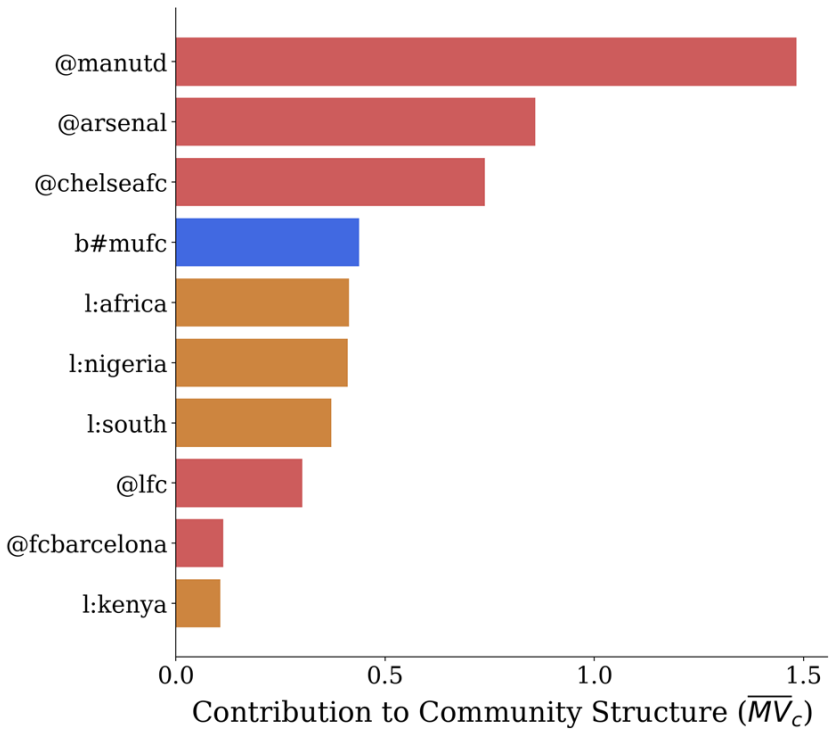

(a)

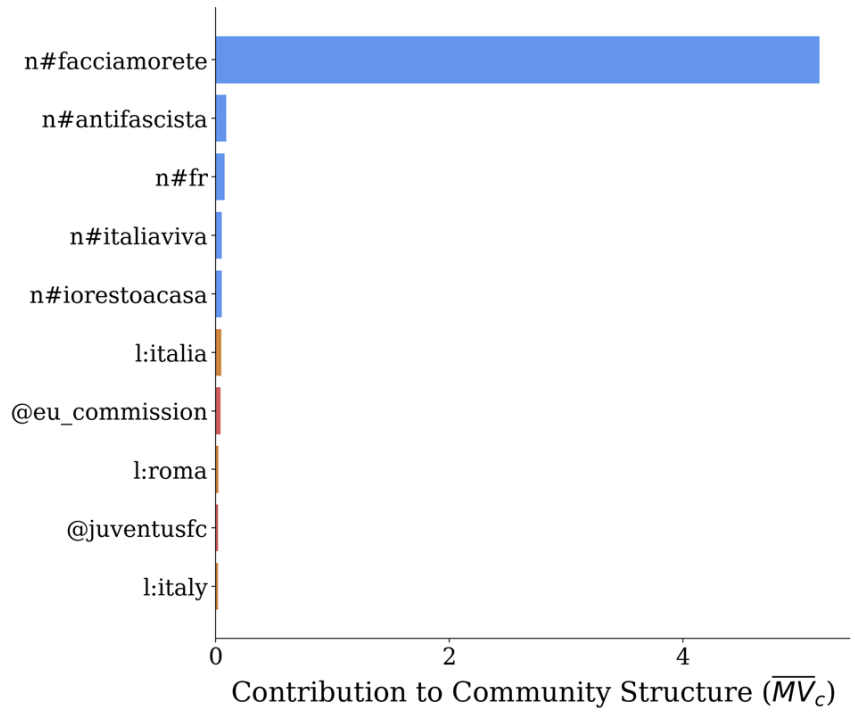

(b)

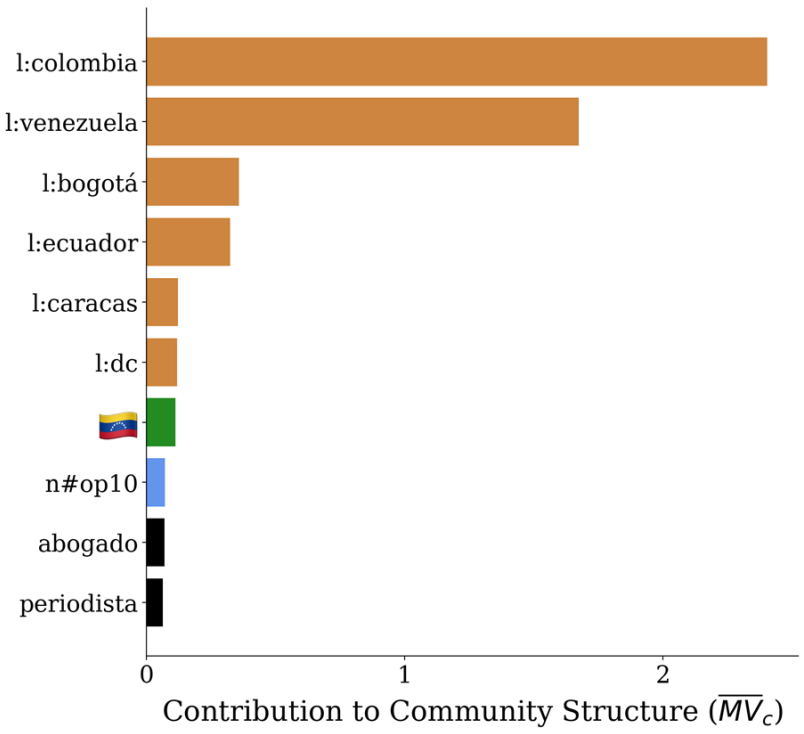

(c)

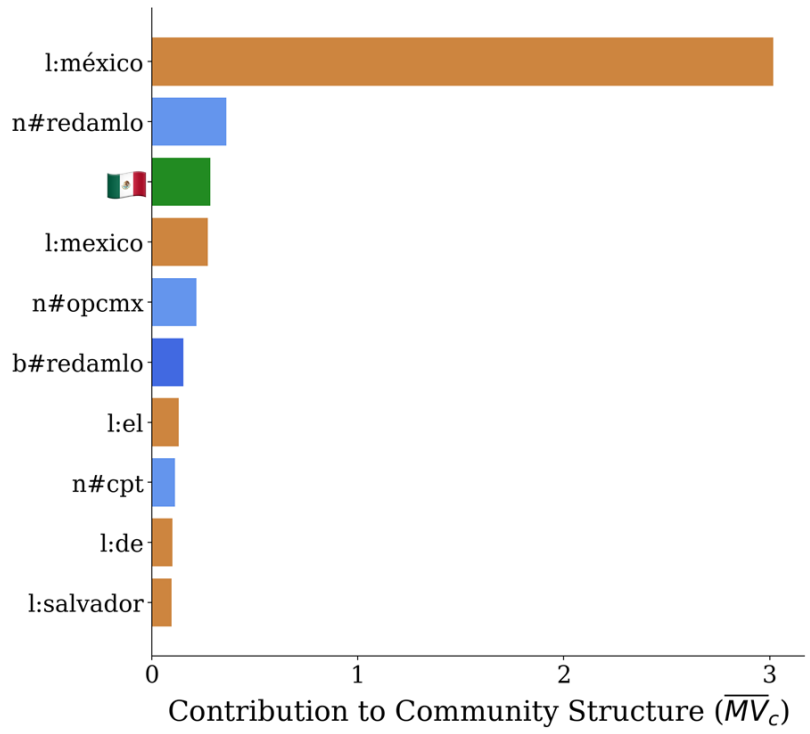

(d)

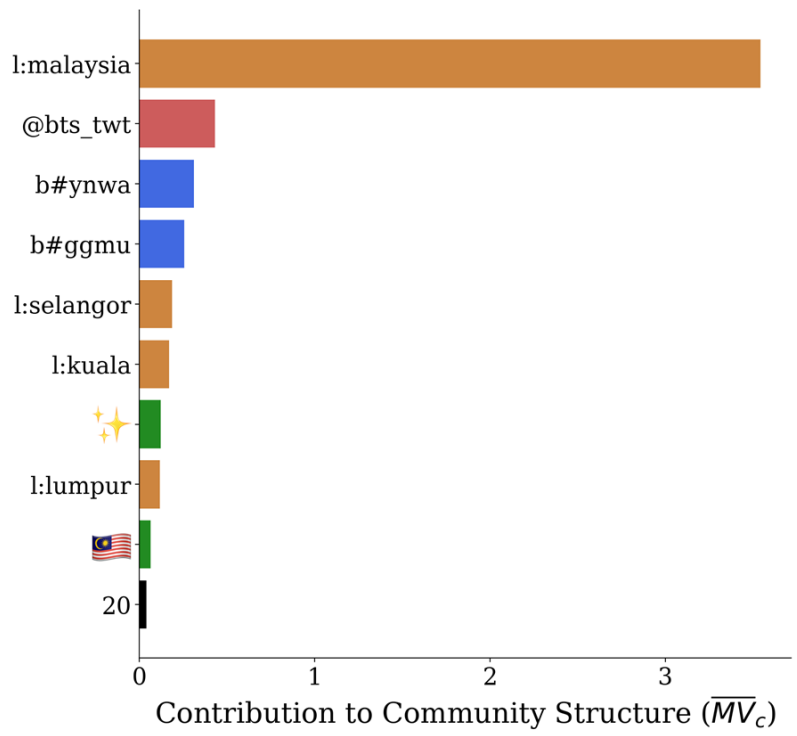

(e)

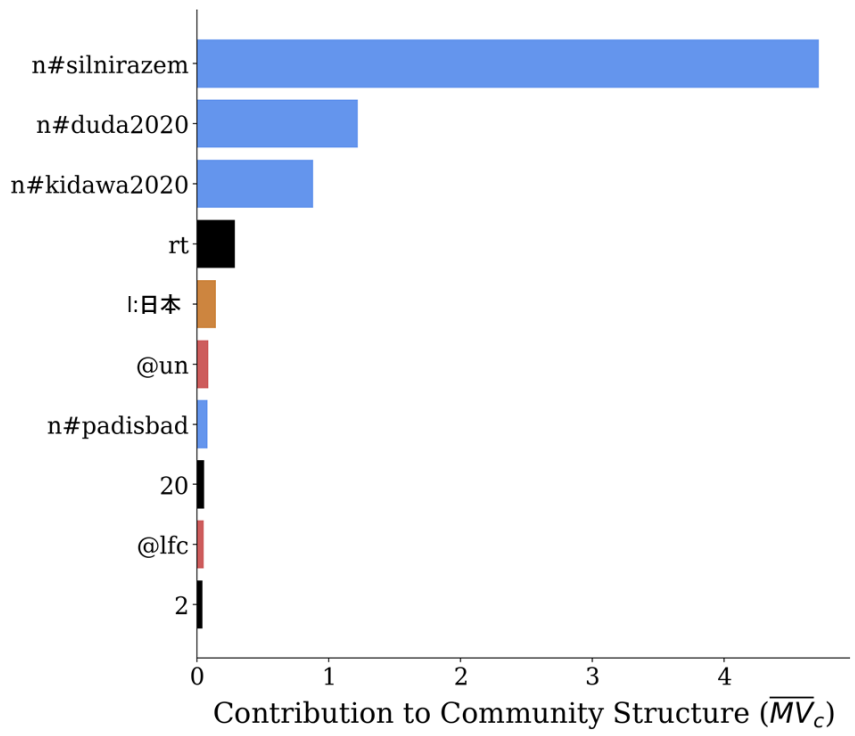

(f)

**Fig. B11**: Prototypes of the communities 13-18 in the *COVID* dataset.

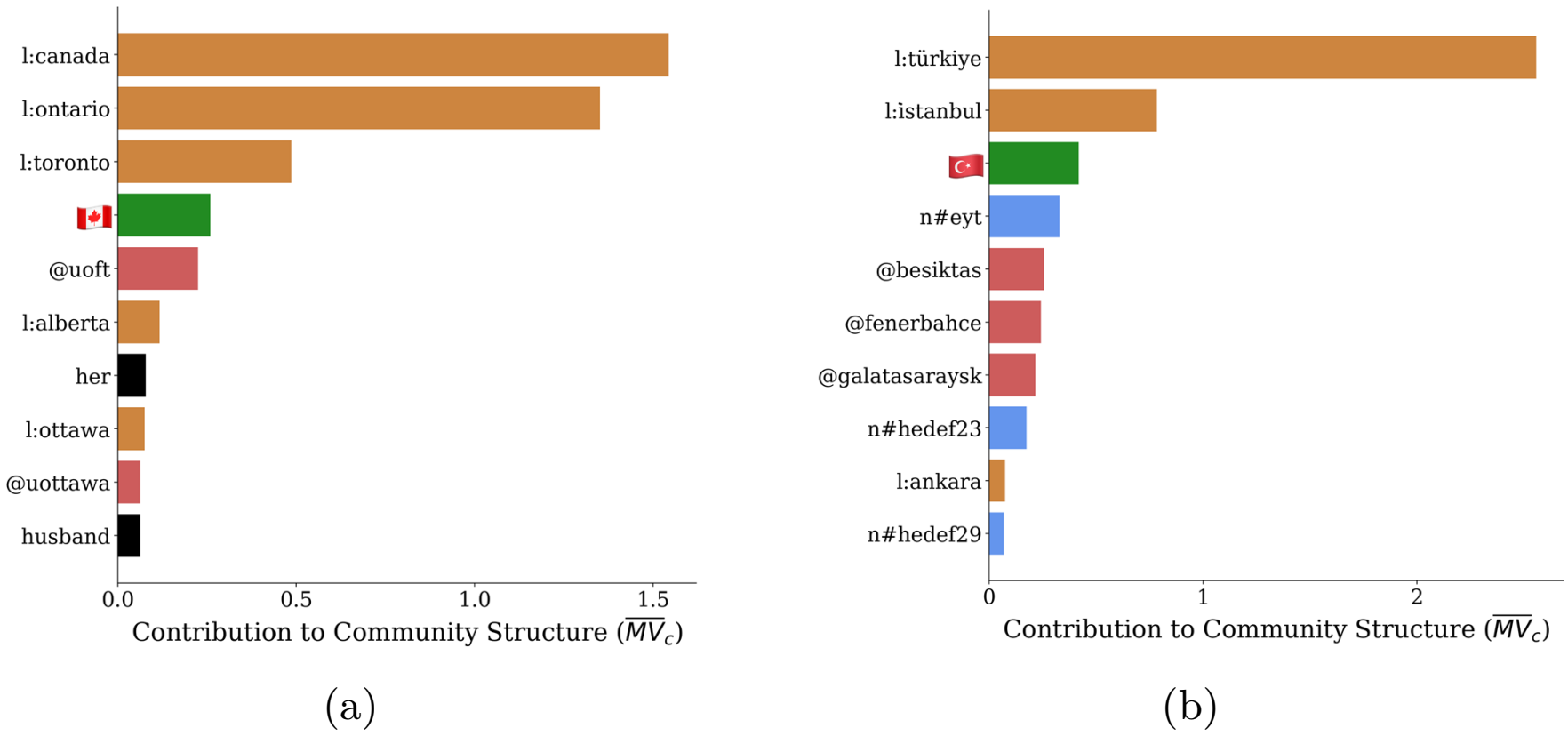

**Fig. B12**: Prototypes of the communities 19-20 in the *COVID* dataset.

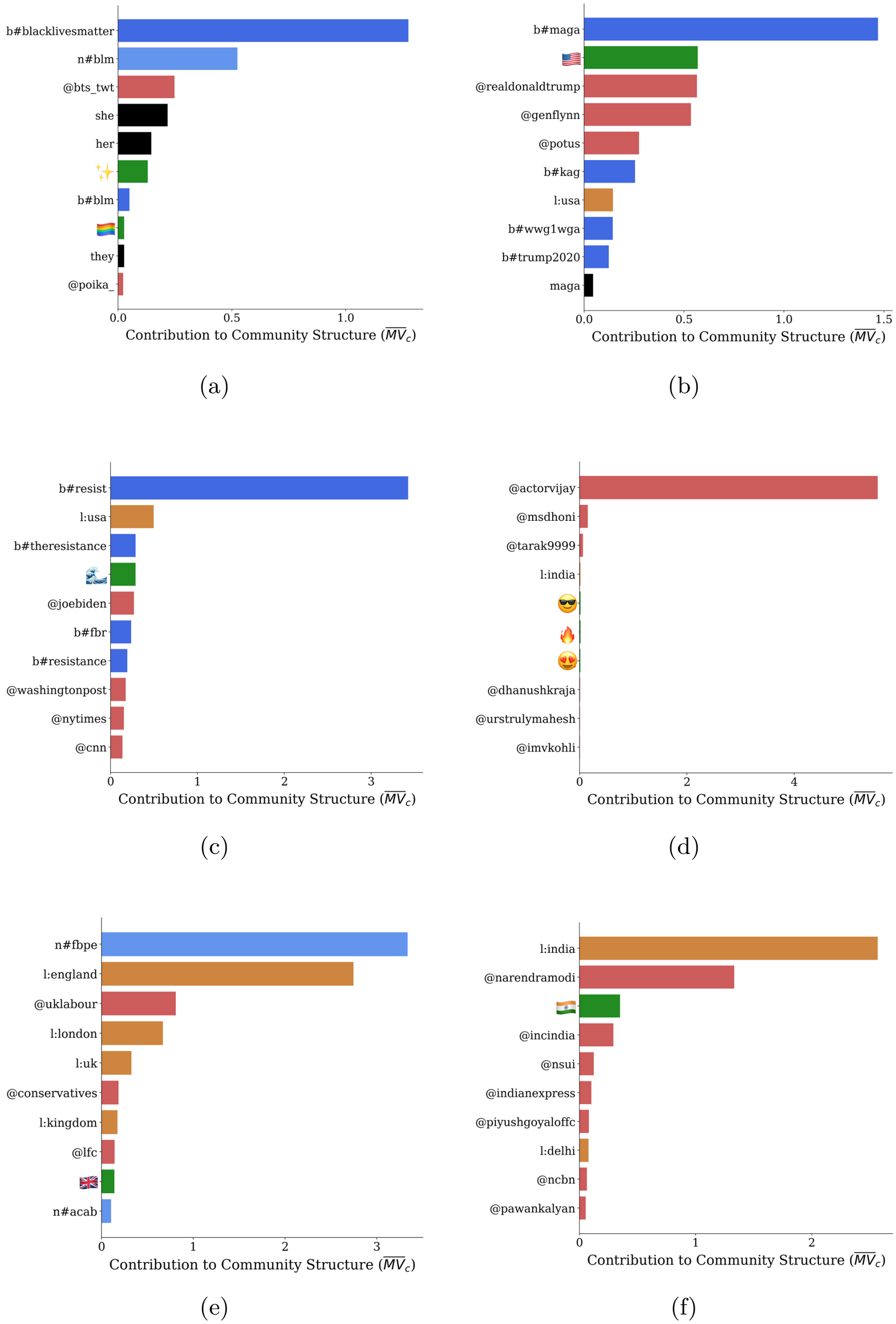

**Fig. B13**: Prototypes of the communities 1-6 in the Reopen dataset.

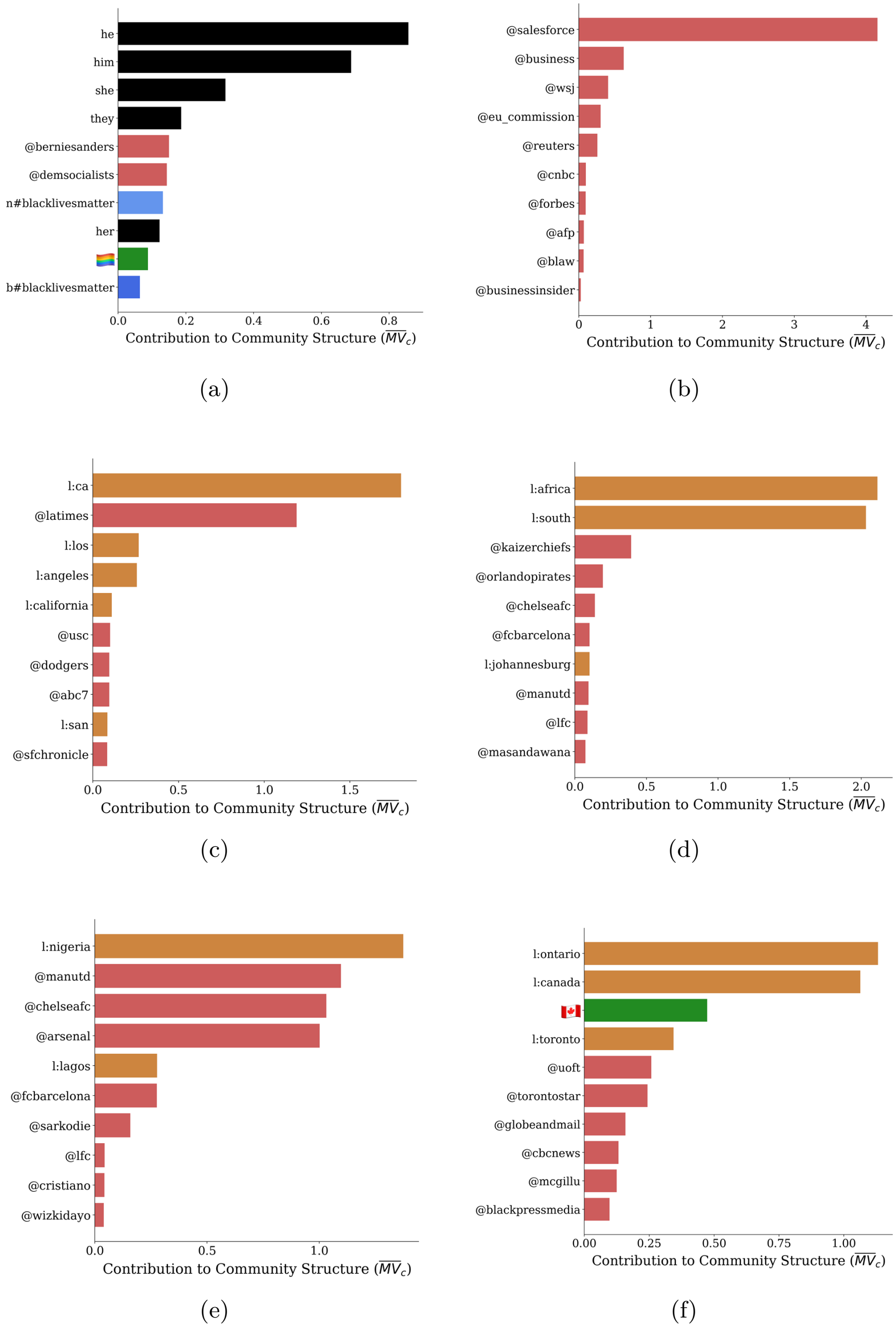

**Fig. B14**: Prototypes of the communities 7-12 in the Reopen dataset.

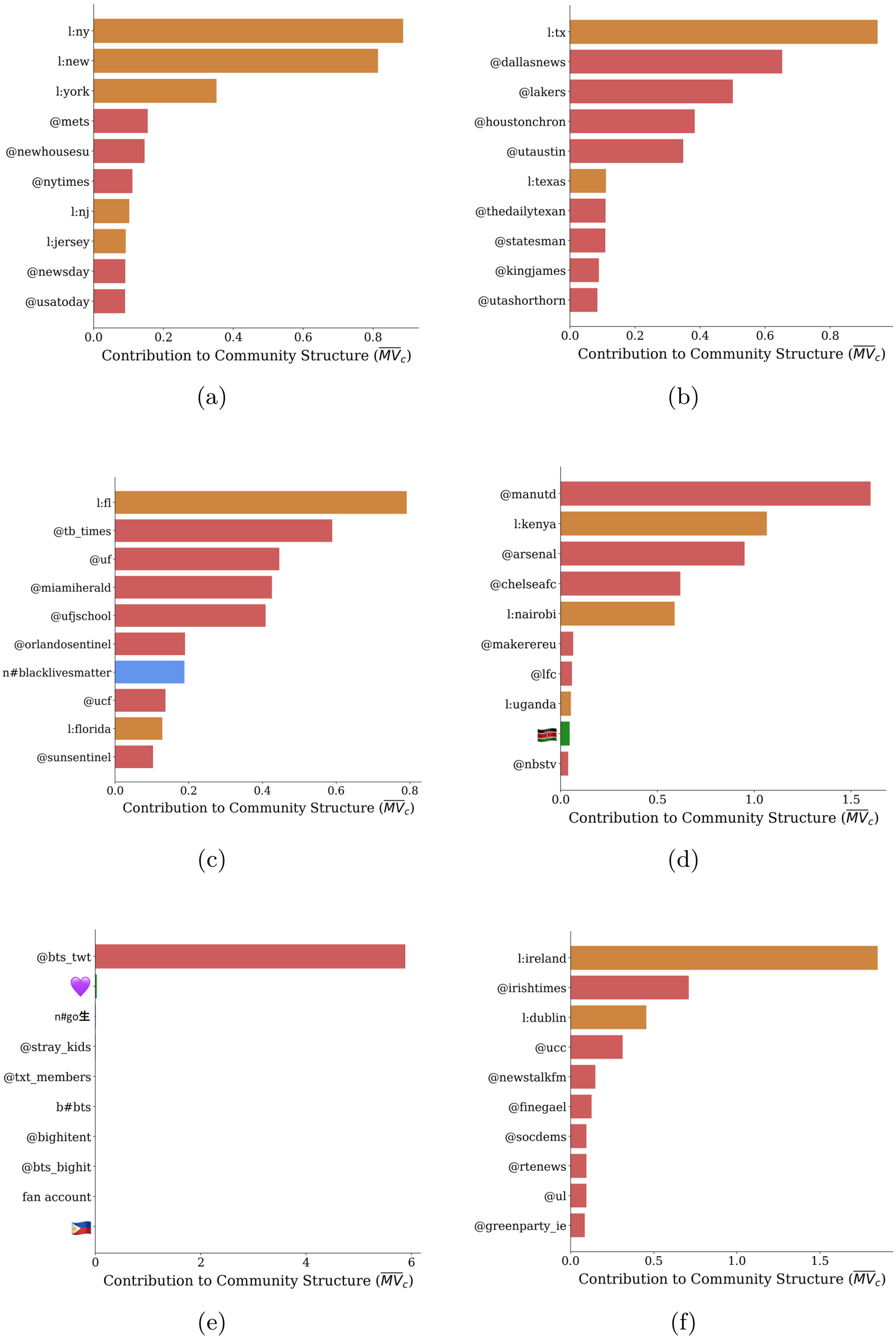

**Fig. B15**: Prototypes of the communities 13-18 in the Reopen dataset.

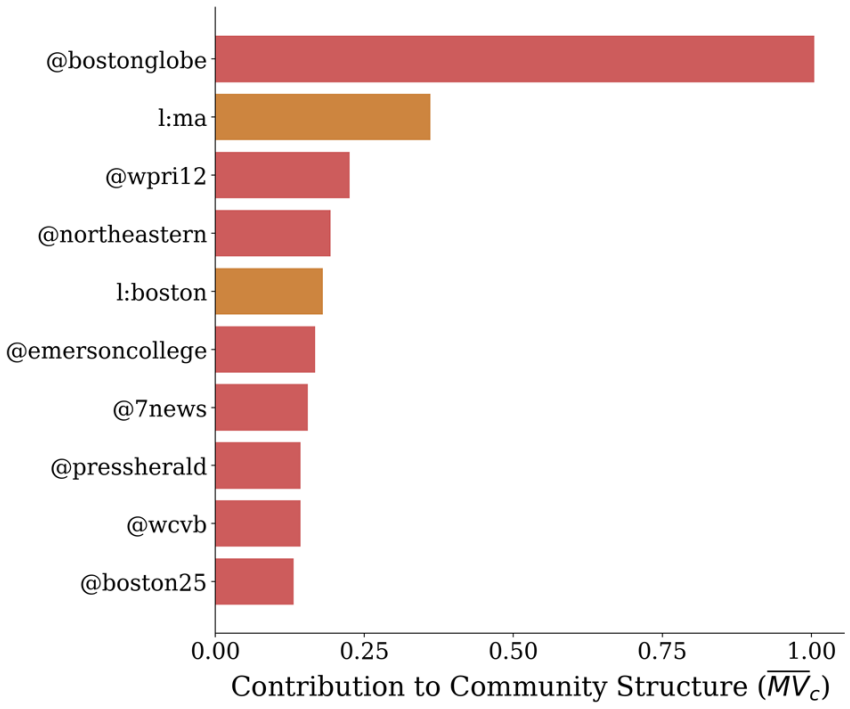

(a)

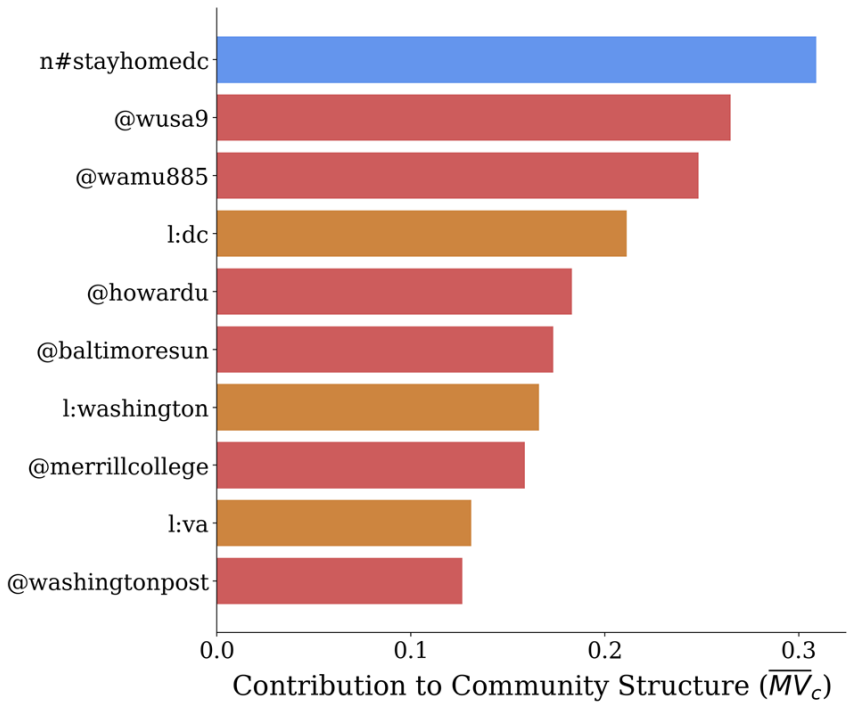

(b)

**Fig. B16**: Prototypes of the communities 19-20 in the Reopen dataset.